\documentclass[journal]{IEEEtran}

\usepackage{cite}      
\usepackage{pgf}
\usepackage{graphicx}  
 \DeclareGraphicsExtensions{.eps,.jpg,.png,.pnm}
\begin{document}
\title{Evolution in Materio: Exploiting the Physics of Materials for Computation}
\author{Simon L. Harding, Julian F. Miller and Edward A. Rietman}
\maketitle

\markboth{ Evolution in Materio: Exploiting the Physics of Materials for Computation}{Shell \MakeLowercase{\textit{et al.}}: Evolution in Materio: Exploiting the Physics of Materials for Computation}

\pagenumbering{arabic} \setcounter{page}{1}

\parskip 4.0mm   

\begin{abstract}
We describe several techniques for using bulk matter for special purpose computation. In each case it is necessary to use an evolutionary algorithm to program the substrate on which the computation is to take place. In addition, the computation comes about as a result of nearest neighbour interactions at the nano- micro- and meso-scale. In our first example we describe evolving a saw-tooth oscillator in a CMOS substrate. In the second example we demonstrate the evolution of a tone discriminator by exploiting the physics of liquid crystals. In the third example we outline using a simulated magnetic quantum dot array and an evolutionary algorithm to develop a pattern matching circuit. Another example we describe exploits the micro-scale physics of charge density waves in crystal lattices. We show that vastly different resistance values can be achieved and controlled in local regions to essentially construct a programmable array of coupled micro-scale quasiperiodic oscillators. Lastly we show an example where evolutionary algorithms could be used to control density modulations, and therefore refractive index modulations, in a fluid for optical computing.
\end{abstract}

\section{Introduction}

There are many physical processes that can be described as a computation. For example, crystal growth from nucleation, corrosion-dendrites on an electrochemical electrode, a drop of ink dispersing in a glass of water, are all physical/chemical processes of increasing complexity that can be thought of as a computation. Further, since biological systems are part of the physical universe, the development of an organism from a fertilized egg is also a computational process. The common element in each of these processes is the fact that the computation is taking place only between nearest neighbors. There is no global clock, or central processor to distribute tasks to the individual processes comprising the overall system. 

Many of these processes are either difficult or way beyond current computational abilities for modelling.  Given a supercomputer, and the set of differential equations and boundary conditions that describe some of these processes, it is likely that we would find that our computed results are only an approximation of the real-world system. The world is a better model of itself than the models we can induce from our data.  Many of these problems are not only computationally intractable, but also computationally undecidable \cite{Wolfram:1985,Moore:1990}. 

As an example, consider an array of magnetic spins in which each site takes on only one of two spin states. At a high temperature the spins will be randomized, but as we cool the array down to much lower temperatures we will find spatial correlations among the spins. This system is computationally tractable only if we make certain simplifying assumptions. But even then, we cannot compute the exact spatial correlations, only the general picture. This example is a particularly interesting problem because we can compute correlations either using detailed quantum mechanics and differential equations, or we can utilize automata theory and obtain essentially the same result (more on this later). Of course the automata theory approach is computationally much faster then the differential equation approach, and the real-world process is even faster. Feynman has implied that the automata theory approach is a potentially more realistic description of the dynamics at the meso-, micro- and nano-scale, than systems of differential equations. The tiny ``computational agents'' at those scales do not compute differential equations. They simply interact with their nearest neighbors and swap information, as Feynman says\cite{feynman67}:


\begin{quote}
``It always bothers me that according to the laws [of physics] as we understand them today, it takes a computing machine an infinite number of logical operations to figure out what goes on in no matter how tiny a region of space and no matter how tiny a region of time. How can all that be going on in that tiny space? Why should it take an infinite amount of logic to figure out what one tiny piece of space-time is going to do?'' 
\end{quote}

If the only processes taking place is information being swapped by nearest neighbors, then as Stephen Wolfram proposes, there may be a universal rule set that governs nearly all the dynamics observed in the universe at all scales \cite{Wolfram:2002}.
 
In this manuscript we will describe methods of exploiting the nearest neighbor physics of meso-scale phenomena for computation. As  Yashihito \cite{Yashihito:1994}, and many others have pointed out, as the device sizes shrink and the level of integration in microcircuits increases we will more closely approach the nanoscale. What was clearly articulated by Yashihito is that we should be able to use matter itself for our computations. We should be able to exploit the molecular dynamics and meso-scale physics for computations. Yashihito did not make explicit suggestions on how to undertake this task.  Miller suggested a variety of physical systems that might be configured to carry out computation\cite{Miller:2002}. One of the suggestions was liquid crystal. This has recently been shown to be possible; Harding and Miller \cite{harding04a,harding04b} have demonstrated for the first time that liquid crystal can be evolved to do analogue filtering. This will be discussed in section \ref{Evolution In Materio}. In the following sections we outline a number of approaches to exploit the physics of materials for computation. We describe numerical and/or empirical results for our suggestions.

\begin{figure}
\centering
\includegraphics[width=50mm]{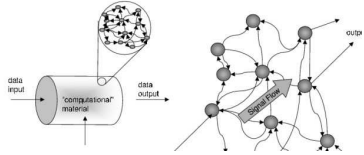}
\caption{Schematic of proposed computational system with bulk matter}
\label{fig:computationalbulkmatter}
\end{figure}

Figure \ref{fig:computationalbulkmatter} shows a schematic of the proposed technology. Basically, we will utilize a block of matter (solid, liquid, or gas) that allows us to change its properties/behavior by external forces. The external forces induce property/ changes, which we can consider to be a sort of ``computer program.'', so that there is a direct link between the external forces that we have control over and the induced changes in the block of matter. By measuring the behaviour of the altered block of matter, we can essentially submit input data to the sample and receive output data. In this way we have performed a type of computation. 

Of course we cannot directly program the molecular dynamics and we do not have control of the molecules, at least not directly. The molecules will interact with their nearest neighbor and we can exploit these phenomena along with the state changes in regions of the block of matter, in order to perform computations. Figure \ref{fig:computationalbulkmatter} shows a nanoscale schematic.

In the following sections we introduce the concept of programmable matter by first describing a well-known "programmable surface" exploited by electrical engineers. Then we describe several examples of programmable matter. Our first example describes a technique that can be used to "program" liquid crystals to perform a computation in the form of signal processing. Following that we describe the use of a magnetic quantum dot array for associative memory. In the next example we introduce the idea of a programmable Fermi surface in certain types of solid-state crystal lattices, in which we describe a prototype system and present some simulation results. In a final example we describe a technique to use acoustic pulses to modulate the density of bulk matter and thereby modulate the refractive index of the material. These refractive index modulations could be used for optical computing.

We stress that some of the examples presented are in preliminary stages of investigation. Therefore, the paper is somewhat speculative but we believe the early results tend to support our speculations, and we believe the preliminary results will be of interest to a wider research community.

\section{Field Programmable Gate Arrays}

\begin{figure}
\centering
\includegraphics[height=\columnwidth, angle=270]{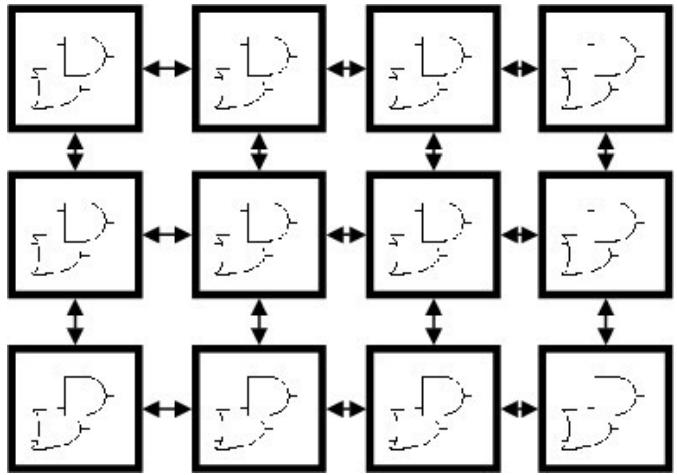}
\caption{Schematic of the architecture for an FPGA.  Each cell can behave like any two-input, one-output Boolean logic function.}
\label{fig:FPGAarchitecture}
\end{figure}

Electrical and computer engineers utilize a chip known as a Field Programmable Gate Array (FPGA). These chips are capable of being configured with software and can emulate many types of digital circuits. The architecture, schematically shown in Figure \ref{fig:FPGAarchitecture}, is analogous to how we can exploit the physics of materials for computation, so a detailed discussion is important. 

The chip consists of an array of programmable logic cells that are connected to their nearest-neighbors. Each cell can exhibit any two-input one-output Boolean logic function.


In addition the connections between the cells are programmable so essentially any Boolean network can be configured within the space limitations of the chip. The chip essentially is a sea of gates - a programmable surface. 

Languages exist for programming these chips. In practice one develops a ``circuit'' which is then downloaded into static random access memory (SRAM). When a system utilizing this combination of SRAM and FPGA is powered up the SRAM sends configuration information to the FPGA and effectively ``wires'' up the logic cells. This download or wiring can take several milliseconds depending on the complexity of the circuit and the number of bits being downloaded.

It is also possible to program the FPGAs as if they were black-boxes without any knowledge of the internal connections. This knowledge-free heuristic programming approach is exactly the same technique we can use to exploit the physics of materials for computation. The technique is called evolutionary programming. 


Wolfram was one of the first to suggest using evolutionary algorithms for evolving real-world systems that may be too complex for humans to engineer\cite{Wolfram:1986}. 

One of the first hardware implementations using evolutionary programming in an FPGA was described by \cite{Thompson:1996} and \cite{Thompson:1997}. They describe evolution of an analog filter in an FPGA. Significantly, they discovered that evolution designed circuits that were often irrational from an engineering design perspective. The circuits would exploit parasitic capacitances and inductances of the transistors. Thompson was the first to point out that evolution was exploiting the physics of the gate in order to achieve the desired fitness. Later work by him \cite{Thompson:2000} was on evolving circuits (in simulation) with single electron transistors. These transistors are expected to operate at 0 K, and so the simulations typically assume this temperature. Instead Thompson and Wasshuber assumed 340mK, which is too hot for the single electron transistor, and thermal noise becomes significant. What they found was that evolution was able to exploit the thermal noise to achieve the goal of building a NOR gate. Later simulation work by Thompson describes the evolution of built in self-test circuits for evaluating sequential and combinatorial circuits\cite{Garvie:2003}. This was unique in that the self-test circuit actually utilized some of the same components used by the main circuit. Finally \cite{Raichman:2003} describe evolution in FPGA circuits. They describe a tone discriminator very similar to \cite{Thompson:1997}. Harding and Miller \cite{harding04a,harding04b} have recently shown that a similar technique can be used to exploit the molecular interactions in liquid crystal and they have evolved ``circuits'' that perform analog filtering(see section \ref{Evolution In Materio}).

In the following,  we review some work of Huelsbergen et al concerning the utilisation of a genetic algorithm and an FPGA to evolve an oscillator circuit\cite{Huelsbergen:1998,Huelsbergen:1999}. In this work, they selected a region of the chip to be exploited, and then decided which pin would be the output where the oscillations could be observed. Recall that the cells and the connections can be configured. The genetic algorithm (GA) - a type of evolutionary algorithm - is easily described. In order to use GA to find an oscillator on this programmable surface (the sea of gates), the configuration information is coded as a long bit string of many thousands of bits. This string is called the chromosome or the genome. Different segments of the string (genes) code for specific configuration information in the FPGA. For example, one segment could configure the cells and another segment could configure the connections. In practice it is a little more complicated, in as much as different segments of the string would code for north connections, another segment of the string would code for the south connections, etc. The genome consisted of about 10,000 bits.

Since we know that the genome codes all the information to be sent to the FPGA and represents the entire configuration of the circuit, we then need some fitness metric to evaluate the said circuit. Huelsbergen et al. chose to monitor the pulses (or lack of pulses) on a selected output pin. But they needed to monitor the pin long enough to integrate over that time period, to determine if the circuit was oscillating, and determine the frequency of the oscillations.

\begin{figure}
\centering
\includegraphics[height=\columnwidth,angle=270]{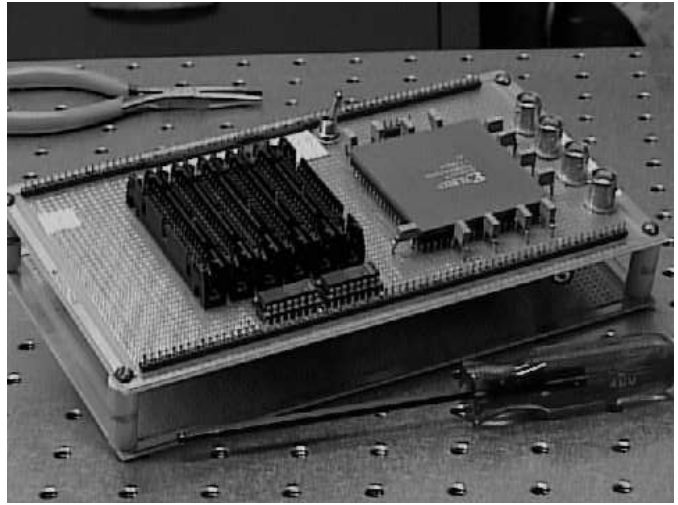}
\caption{FPGA - programmable surface - used by \cite{Huelsbergen:1998,Huelsbergen:1999} for their experiments in evolution in a physical realm.}
\label{fig:FPGA_programmable_surface}
\end{figure}

After selecting the genome - the bit string representing the circuit - and fitness - a metric of how well the circuit behaves - the next step in the GA consists of initializing a population of these strings and evaluating each of them. This consists of actually downloading each string in sequence and evaluating how well the circuit behaves with respect to the goal. The worst performing strings are discarded. The best strings are kept and segments between best performing strings are swapped with each other to create a new population of strings, which is then evaluated. The process repeats and eventually converges to strings/circuits that have either the desired behavior or is close to the desired behavior. Figure \ref{fig:FPGA_programmable_surface} is a photo of the circuit they used for their experiments.

\begin{figure}
\centering
\includegraphics[height=\columnwidth, angle=270]{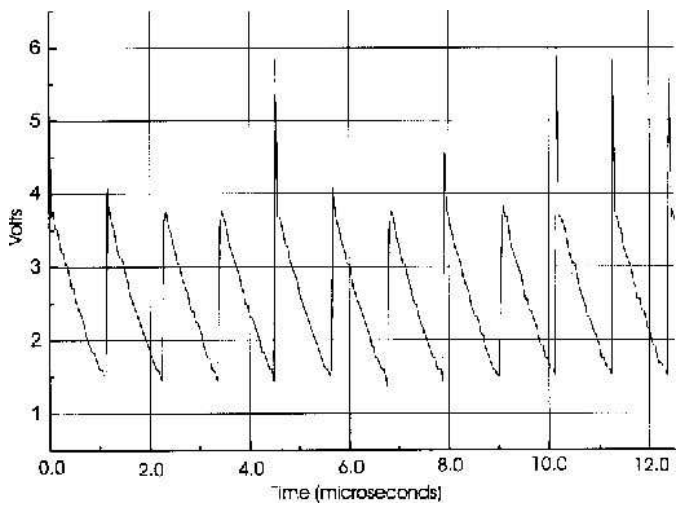}
\caption{Oscilloscope trace from evolved oscillator circuit in CMOS hardware (Huelsbergen et al. 1998)\cite{Huelsbergen:1998}.}
\label{fig:Oscilloscope_trace_from_evolved_oscillator_circuit}
\end{figure}

There was some peculiarity in their results, which have impact on the use of evolution algorithms for configuring physical media. Figure \ref{fig:Oscilloscope_trace_from_evolved_oscillator_circuit} is an example of the oscilloscope trace from one of the oscillators that was evolved. Firstly, notice that the waveform is a saw-tooth, not the expected square wave from a CMOS device. Secondly, notice the voltage levels. The chip can be configured at TTL or CMOS voltage levels. In this experiment it was configured at TTL and this is seen in the voltage levels for saw-tooth wave. The lowest voltage point is about 1.5 volts. This is the highest voltage that TTL logic will register logic low. The highest voltage, excluding spikes, is at 3.8 volts. This is the lowest that TTL logic will register high logic. This, along with the random voltage spikes above 5.0 volts indicates the transistors are operating in the linear regime and relaxing by releasing random spikes. This bizarre behavior from CMOS logic is due to the fact that there were no constraints in the fitness function to force the logic to operate in the digital regime. Evolution does not know anything about the physical medium. It is blindly optimizing a fitness function given the physical medium. So the evolved circuit essentially exploits the physics of the transistors themselves without any specific instructions to do so. Furthermore, the evolved circuit is sensitive to the actual FPGA. If we replace the FPGA with another chip and download the evolved circuit it will behave slightly differently with respect to the frequency. This is further evidence that evolution is exploiting the physics of the transistors. When we replace the transistors with a different physical set of transistors (i.e. replace the silicon chip) we get different behavior.

In summary, we have shown that we can use evolution to design computational circuits with a ``sea of gates.'' Evolution will exploit the substrate at the lowest level possible to accomplish the desired task. Evolution will actually exploit the physics of the gates and the variation in the silicon crystals to achieve its goals. There are several implications for exploiting the physics of matter. Firstly, we do not need to pre-compute anything. Secondly, by definition, since we are manipulating physical matter with the evolutionary algorithm if a solution can be found then we will find it and it will be physically realizable. Thirdly, evolution will cobble together any thing to get the problem solved. It will not necessarily ``build'' the ``computational system'' we expect, because it will exploit the physics of the medium in ways we may not even imagine.

\section{Programming a Liquid Crystal Substrate}
\label{Evolution In Materio}
\subsection{Introduction}
%
As shown above in Figure \ref{fig:Oscilloscope_trace_from_evolved_oscillator_circuit} and discussed by Thompson et al. \cite{Thompson} and Huelsbergen et al \cite{Huelsbergen:1998,Huelsbergen:1999} it is now obvious that evolutionary algorithms exploited subtle physical properties of the FPGA and associated circuits in order to solve a particular problem. It is not fully understood what properties of the FPGA substrate were being exploited. This lack of knowledge of how the system works prevents engineering the design of systems to exploit the complex physical characteristics. We argue that the lesson that should be drawn from the work of Thompson is that evolution may be used to exploit the properties of a wider range of materials than silicon. This further suggests an exploration through artificial evolution. We refer to this as ``evolution in materio.'' Miller suggested a good candidate for evolution in materio would be liquid crystals \cite{Miller02}. Recently this suggestion has been supported by work of Harding and Miller \cite{harding04b}, who showed that it is relatively easy to configure (using computer controlled evolution) liquid crystals to perform various forms of computation.  

For this work we have chosen to use a genetic algorithm, as there is an already well established field of evolvable hardware from which we can draw experience. Other search methods have been suggested for material systems, and will probably work to the same level.  In previous work, Toffoli argues that simulated annealing may provide a more suitable programming technique for programmable materials\cite{Toffoli99}. This technique shares many similarities with evolutionary algorithms, however, simulating annealing can perform less efficiently as it does not use a population based approach and therefore can more easily become trapped in local attractors.
  
\subsection{Liquid Crystal}
Liquid crystal (LC) is commonly defined as a substance that can exist in a mesomorphic state \cite{Demus98}\cite{Khoo95}.  Mesomorphic states have a degree of molecular order that lies between that of a solid crystal (long-range positional and orientational) and a liquid, glass or amorphous solid (no long-range order). In LC there is long-range orientational order but no long-range positional order. 

Aromatic LC is often called a benzene derivative. There is also heterocyclic LC where one or more of the benzene rings are replaced with pyridine, pyrimidine or other similar group. LC can also have metallic atoms (as a terminal group) in which case they are called organometallic compounds. Chemical stability is strongly influenced by the linkage group. Compounds where the aromatic rings are directly linked are extremely stable. LC tends to be transparent in the visible and near infrared and quite absorptive in UV. 
\\
There are three distinct types of LC: lyotropic, polymeric and thermotropic. Thermotropic LC (TLC) is the most common form and is widely used. TLC exhibit various liquid crystalline phases as a function of temperature. They can be depicted as rod-like molecules and interact with each other in distinctive ordered structures. TLC exists in three main forms: nematic, cholesteric and smectic. In nematic LC, the molecules are positionally arranged at random, but all share a common alignment axis. Cholesteric LC (or chiral nematic) is like nematic however they have a chiral orientation. In smectic LC, there is typically a layered, positionally disordered structure. In type A the molecules are oriented in alignment with the natural physical axes (i.e normal to the glass container, depicted by the arrow), however in type C, the common molecular axes of orientation is at an angle to the container. 

There is a vast range of different types of liquid crystal. LC of different types can be mixed. LC can be doped (as in Dye-Doped LC) to alter their light absorption characteristics. Dye-Doped LC film has been made that is optically addressable and can undergo very large changes in refractive index \cite{Khoo98}. There are Polymer-Dispersed Liquid Crystals these can have tailored electrically controlled light refractive properties. Another interesting form of LC being actively investigated is Discotic LC. These have the form of disordered stacks ( 1-dimensional fluids) of disc-shaped molecules on a two dimensional lattice. Although discotic LC is an electrical insulator, it can be made to conduct by doping with oxidants \cite{Chandrasekhar98}. LC is widely known as useful in electronic displays, however, there are in fact, many non-display applications too. There are many applications of LC to electrically controlled light modulation: phase modulation, optical correlation, optical interconnects and switches, wavelength filters and optical neural networks. In the latter case, LC is used to encode the weights in a neural network \cite{Crossland98}.

\begin{figure}[ht]
	\centering
		\includegraphics[width=\columnwidth]{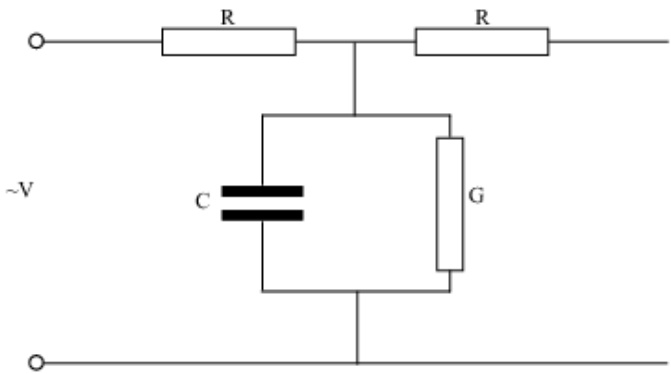}
	\caption{Equivalent circuit for LC}
	\label{fig:lc-elec-equiv}
\end{figure}

Figure \ref{fig:lc-elec-equiv} shows the equivalent electrical circuit for liquid crystal between two electrodes when an AC voltage is applied.  The distributed resistors, R, are produced by the electrodes.  The capacitance, C, and the conductance, G, are produced by the liquid crystal layer \cite{naumov:98}. 

\subsection{An Evolvable Motherboard with a FPMA}


We have been experimenting with a hardware system that enables programmability and reconfiguration through a semiconductor cross point array switch. We refer to this system as a evolvable motherboard (EM) \cite{Layzell1998}. When the EM is connected to a material substrate (e.g. liquid crystal display) or discrete electronic components \cite{Layzell1998, Crooks02} we refer to the system as a field programmable matter array (FPMA). The EM is connected to a PC that is used to control the evolutionary processes. The EM also has digital and analog I/O that can be accessed for test and recording of the response of the material under evolution.

\begin{figure*}
\centering
\includegraphics[width=\columnwidth,angle=270]{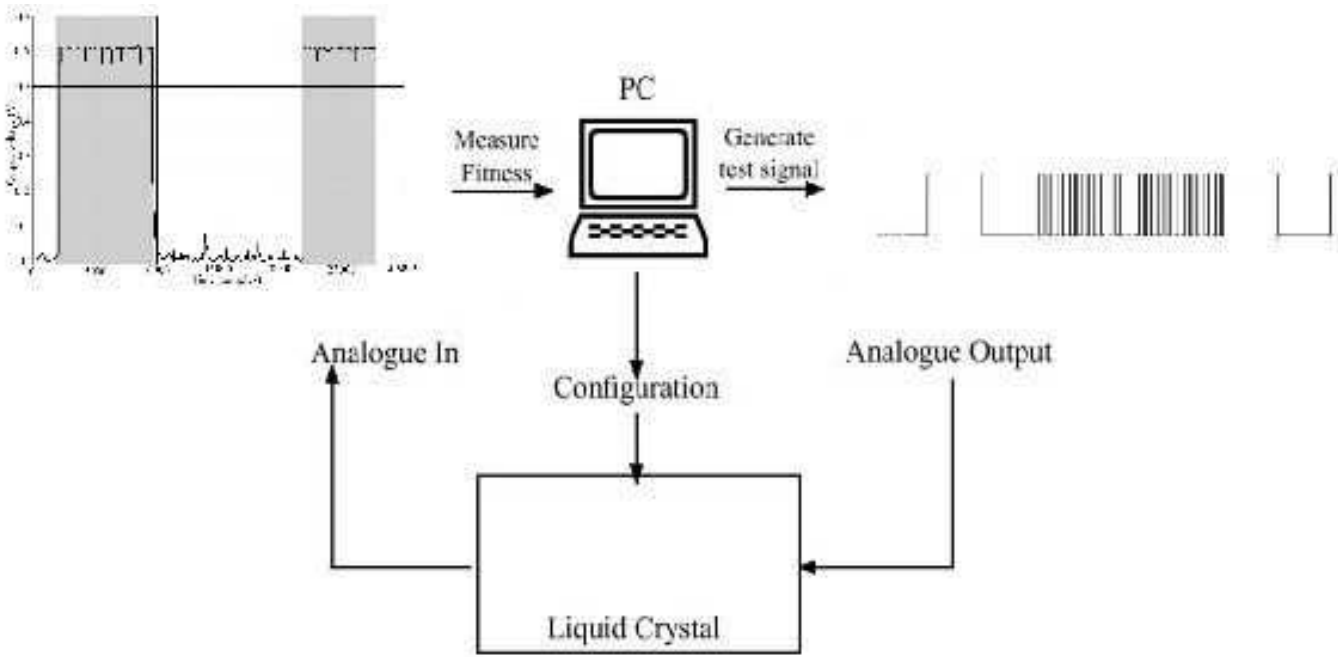}
\caption{Equipment configuration}
\label{fig:evolvatronsetup}
\end{figure*}

In the experiments presented here, a standard liquid crystal display with twisted nematic liquid crystals was used as the medium for evolution. It is assumed that the electrodes are indium tin oxide. Typically, such a display would be connected to a driver circuit.  The driver circuit has a configuration bus on which commands can be given for writing text or individually addressing pixels so that images can be displayed.  The driver circuit has a large number of outputs that connect to the wires on the matrix display. When displaying an image, appropriate connections are held high, at a fixed voltage - the outputs are typically either fully on or fully off.

\begin{figure}
\centering
\includegraphics[width=\columnwidth]{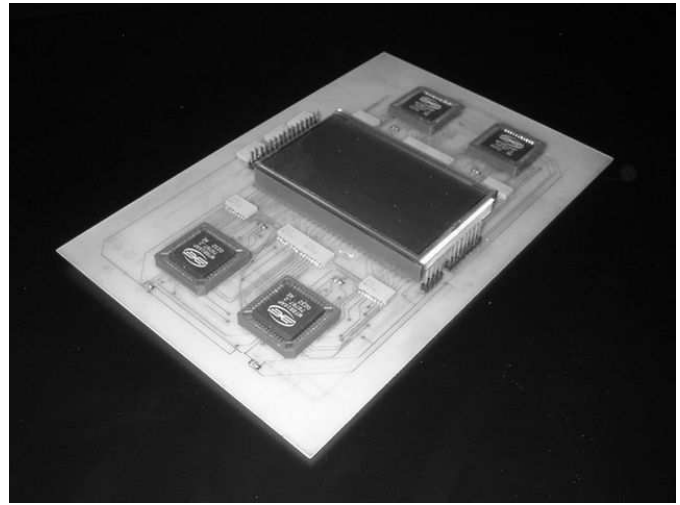}
\caption{Photograph of the LCEM prototype circuit.}
\label{fig:lcem}
\end{figure}

Such a driver circuit was unsuitable for the task of intrinsic evolution.  There is a need to be able to apply both control signals and incident signals to the display, and also record the response from a particular connector. Evolution should be allowed to determine the correct voltages to apply, and may choose to apply several different values.  The evolutionary algorithm should also be able to select suitable positions to apply and record values.  A standard driver circuit would be unable to do this satisfactorily.  Hence a variation of the evolvable motherboard was developed in order to meet these requirements.  

The Liquid Crystal Evolvable Motherboard (LCEM) is a circuit that uses four cross-switch matrix devices to dynamically configure circuits connecting to the liquid crystal.  The switches are used to wire the 64 connections on the LCD to one of 8 external connections.  The external connections are: input voltages, grounding, signals and connections to measurement devices.  Each of the external connectors can be wired to any of the connections to the LCD.

\begin{figure}
\centering
\includegraphics[width=\columnwidth]{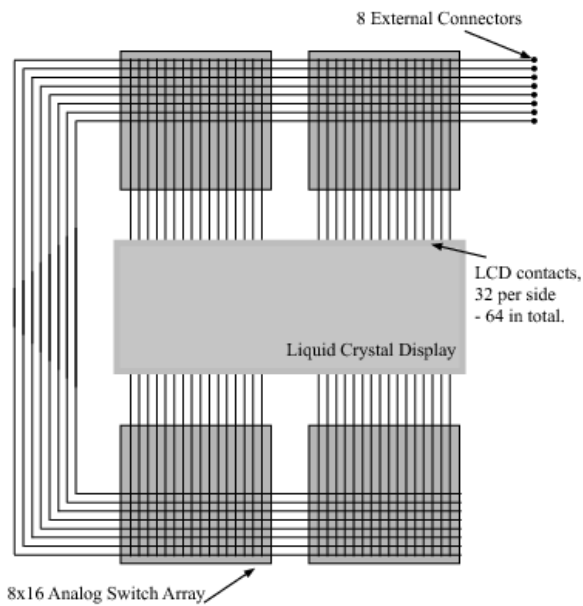}
\caption{Schematic of LCEM}
\label{fig:lcemschem}
\end{figure}

The external connections of the LCEM are connected to the computers analogue inputs and outputs.  One connection was assigned for the incident signal, one for measurement and the other for fixed voltages.  The value of the fixed voltages is determined by the evolutionary algorithm, but is constant throughout each evaluation.  Each of the external connectors can be wired to any of the connections in the LCD (see figures \ref{fig:evolvatronsetup} and \ref{fig:lcemschem}).

In these experiments the liquid crystal glass sandwich was removed from the display controller it was originally mounted on, and placed on the LCEM.  The display has a large number of connections (in excess of 200), however because of PCB manufacturing constraints we are limited in the size of connection we can make, and hence the number of connections.  The LCD is therefore roughly positioned over the pads on the PCB, with many of the PCB pads touching more than 1 of the connectors on the LCD. This means that we are applying configuration voltages to several areas of LC at the same time.

Unfortunately neither the internal structure nor the electrical characteristics of the LCD are known.  This raises the possibility that a configuration may be applied that would damage the device.  The wires inside the LCD are made of an extremely thin material that could easily be burnt out if too much current flows through them.  To guard against this, each connection to the LCD is made through a 4.7Kohm resistor in order to provide protection against short circuits and to help limit the current in the LCD.  The current supplied to the LCD is limited to 100mA.  The software controlling the evolution is also responsible for avoiding configurations that may endanger the device (such as short circuits).

It is important to note that other than the control circuitry for the switch arrays there are no other active components on the motherboard - only analog switches, smoothing capacitors, resistors and the LCD are present.

\subsection{Genetic Representation}

The genetic representation for each individual is made of two parts.  The first part specifies the connectivity; the second part determines the configuration voltages applied to the the LCD.

Each connector on the LCD can be connected to one of the eight external connectors or left to float, figure \ref{fig:lcemschem}.  Each of the connectors is represented by a number from 0 to 7 and no connection is represented by 8.  Hence the genotype for connectivity is a string of 64 integers in the range 0 to 8.

The remainder of the genotype specifies the voltages applied to the pins on the external connector that are not used for signal injection / monitoring. On the LCEM there five such configurable connectors - the other three are connected to ground, the incident signal and to the data recorder.  The voltage is represented as a 16-bit integer, the 65536 possible values map to the voltage levels output from -10V to +10V.  The second section of the genotype is therefore represented as a string of five 16bit integers.

A mutation is defined as randomly taking an element in one part of the genotype and setting it to a randomly selected new value.  Constraints are enforced to prevent illegal configurations.  

We chose not to use genetic recombination as the constraints imposed on this representation would make it difficult to implement and would require many arbitrary decisions to be made on suitable repair techniques.  For example, it is unclear what strategy should be used to fix a genotype where there are multiple outputs and only one is allowed.  For this reason, the evolutionary algorithm used here has no crossover operator. 

In all the following experiments, a population of 40 individuals was used.  The mutation rate was set to 5 mutations per individual.  Elitism was used, with 5 individuals selected from the population going through to the next generation.  Selection was performed using tournament selection based on a sample of 5 individuals.

Evolutionary runs were limited to 100 generations. With each generation taking approximately 60s to evaluate.

\subsection{Evolution Of A Tone Discriminator}
A tone discriminator is a device, which when presented with one of two input signals returns a different response for the each signal. In \cite{Thompson}, on which this experiment is loosely based, the FPGA under investigation was asked to differentiate a 1kHz square wave from a 10kHz square wave, giving a low output for one and a high output for the other.  In this experiment we have arbitrarily chosen two frequencies of 100Hz and 5kHz. Each signal is a square wave, oscillating between 0V and 5V, with equal timing given to the low and high states.  The tones were presented in 250ms bursts with no gap between the tones.  The goal was to evolve a device that would output a low value (\begin{math}\langle\end{math}0.1V) at low frequencies, and high (\begin{math}\rangle\end{math}0.1V) at the higher frequency.  The fitness was calculated as the percentage of samples made where the output was in the correct state for a given input frequency. 

Let S be the vector containing the input sample.  Let $L$ be the length of $S$.  $O$ is a vector containing the output frequency at a given time.  The output frequency can be either HIGH or LOW.  The \begin{math}j\end{math}th element of the set is \begin{math} S\left[j\right] \end{math}. t is the threshold for a low response, i.e. \begin{math}\langle\end{math}0.1V.


\begin{displaymath}
x(i) = \left\{ \begin{array}{ll}
1 & \textrm{if $S\left[i\right]\leq t~and~O\left[i\right]$ = HIGH}\\
1 & \textrm{if $S\left[i\right]\geq t~and~O\left[i\right]$ = LOW}\\
0 & \textrm{otherwise}
\end{array} \right.
\end{displaymath}
\begin{displaymath}
fitness = \frac{\sum_{i=0}^{L} x(i)}{L}
\end{displaymath}

It is important to note that samples taken do not correspond to time as samples are taken on an interrupt, and frequency of sampling may be affected by other processes running on the computer.

\begin{figure}
\centering
\includegraphics[width=\columnwidth]{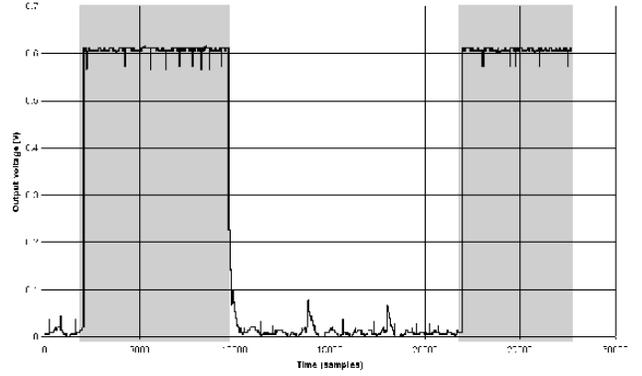}
\caption{Tone discriminator response. Dark areas indicate 5kHz input, light 100Hz}
\label{fig:tonerepsonse1}
\end{figure}

Not all attempts at evolving the discriminator under these conditions were successful, however we did manage to evolve a discriminator with the response shown in Figure \ref{fig:tonerepsonse1}.  Although the output was not stable, there is a clear difference between the behaviour at low and high frequencies. At high frequencies, a high output was obtained for the majority over time and for low frequencies a low output was obtained.

We assume the behaviour stems from capacitative effects originating inside the LCD, and that the system is acting as a form of R-C network.  The crosspoint switches are unlikely to be involved as they are designed for high frequency audio/video signals.  The feed-through capacitance at 1Mhz is 0.2pF and the switch I/O capacitance is 20pf.  This would seem too small to have any filtering effect on these relatively low frequencies.


An interesting observation is that if a configuration is reloaded into the LCEM it fails to work, however if the population containing that solution is allowed to evolve (for another 2 to 3 generations) the behaviour returns.  The cause of this appears to be a lack of stability in the system, however it is unclear if this is caused by the liquid crystal itself or some other component in the system.

\subsection{Summary}
The work described is the first example of the use of computer controlled evolution to build computational processors in liquid crystal. Although liquid crystal appears to be suitable for use as an evolutionary medium, there are many unanswered questions. More work is required to prove that the LC is responsible for the observed results and to attempt to discover how the LC is being exploited.  It may be that the LC layer is being used as some form of configurable continuous RC network - similar to that shown in figure \ref{fig:lc-elec-equiv} but with much richer properties. 

\section{Magnetic Quantum Dots}

There are significant parallels between neural networks, chaotic computing, optical computing, molecular computing, and of course, quantum computing \cite{Calude:1998, GromB:1998, Siegelmann:1999,Sipper:1997, Mange:1998, Sienko:2003}.  All of these have in common the fact that they are ``unconventional models of computation.'' Two of the most interesting computational paradigms that join several of the other unconventional approaches are - quantum dot cellular automata and quantum spin computing. 

Benioff first suggested quantum mechanical computation as Turing machine automata on Hamiltonian lattices \cite{Benioff:1980,Benioff:1982}.  Albert suggested a similar concept of quantum mechanical automata\cite{Albert:1983}. These really significant works no doubt inspired the suggestion of cellular automata for nanometer-scale computing\cite{Biafore:1994}, and inspired Porod et al. for their work on quantum-dot cellular automata \cite{Porod:1999}. (Cellular automata will be introduced shortly.)

Other quantum computing work has drawn parallels with brains. The most notable in this effort is Penrose \cite{Penrose:1989,Penrose:1994} and the more recent is Satinover \cite{Satinover:2001}. Though these works are largely in the area of philosophical research; it has had some impact on more serious physics research such as Kak\cite{Kak:1992} who wrote about state generators and complex neural memories in neural networks. This early Kak paper resulted in his later work on comparisons with brain dynamics and neural networks \cite{Kak:1995}, which is typical in the field. Finally, Ventura implemented competitive learning in a quantum computing neural network\cite{Ventura:1999} ;  Behrman described a spatial quantum neural computer\cite{Behrman:1999}; and Zak described quantum analog computing\cite{Zak:1999}. Though most of this work is very theoretical it an important contribution. Of a more experimental nature, others have described methods for building  quantum networks\cite{Mahler:1995}. Much of that approach suggests quantum spin systems using quantum dots.

One of the earliest suggestions of quantum dot spin systems was described by Smith\cite{Smith:1990}. Their suggestion was to build arrays of GaAs-based quantum dots. A review of related experimental work is given by Chakraborty \cite{Chakraborty:1999}.  Spin transistors and spin electronics are finally attracting a great deal of attention (\cite{Zorpette:2001,DasSarma:2001,Wolf:2001}).

We propose an experimental prototype and describe simulation results, which are a follow up from some interesting observations by Kirczenow et al\cite{Kirczenow:1993} on the dynamics of quantum hall phenomena in arrays of magnetic quantum dots, and some suggestions of \cite{Matsueda:1999} on quantum computing using dipole-dipole block systems. The work by Kirczenow et al. describes two-dimensional arrays of magnetic dots with larger dots on the edges that are used to couple into the array. They show that if the dot is larger in diameter than in height it will take on only two spin states. In this respect these results are similar to \cite{Cowburn:1999} and \cite{Yu:2002}. In the following we describe some simulations we have done to show that it is possible to use a genetic algorithm to program a magnetic quantum dot computer for use in pattern recognition.

\subsection{Ising Spins}

In this section, the physics of spin glasses or Ising models, follows the presentation by \cite{Bar-Yam:1997}. \cite{Mezard:1986} is a good introduction to the statistical physics of spin glasses. The subject of Ising spins has recently been re-introduced by \cite{Hayes:2000}.

\begin{figure}
\centering
\includegraphics[height=2in, angle=270]{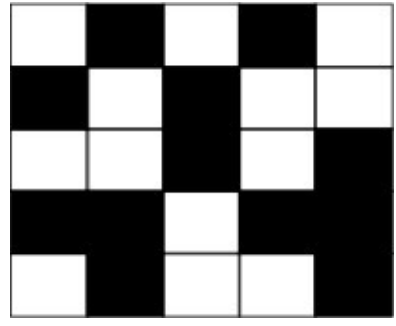}
\caption{Ising spin array for quantum computing}
\label{fig:Isingspinarray}
\end{figure}

Consider a square grid of lattice points. At each point we will place a tiny magnet (a spin) and we will represent this by spin up or spin down. These spins are represented by two colors, black and white, in Figure \ref{fig:Isingspinarray}. Now let the spins flip so the entire system can relax to the lowest energy configuration. This can be done at each temperature and it turns out there is a minimum energy for a given grid of spins at any temperature. At high temperatures the spin arrangement can be highly random with little or no correlations. At lower temperature, we will find that the spins are more spatially correlated. That, in summary, is the entire idea behind Ising models for magnetic spins (a.k.a. spin glasses).

In the above, the energy is a function of the spin, $ E \left[ \left\{ s_i \right\} \right] $, for each spin   in the system. So the total energy is given by

\begin{math}
E \left[ \left\{ s_i \right\} \right] = \sum_i{ e_i(s_i) }
\end{math}
 
where $e_i(s_i)$  is the energy of an individual spin that is not influenced by another neighboring spin. Since our system is binary we can write this equation as

\begin{eqnarray}
\label{ising1}
E \left[ \left\{ s_i \right\} \right] & =  \frac{1}{2} \sum_i{ \left[ e_i(1) - e_i(-1) \right]s_i + } \nonumber  \\
  & \left[ e_i(1) + e_i(-1) \right]  \nonumber  \\
 &  = E_0 - \sum_i{h_is_i}
\end{eqnarray}  

where some obvious terms have been collected together as the ``ground-state'' energy, $E_0$  . The quantity $h_i$  is the energy due to orientation of the spins, and in a magnetic system they represent the external magnetic field.  This external magnetic field will prove to be very useful in programming an Ising spin computer.

In this simple system we can write the probability of a particular configuration of spins by the following equation:\\

\begin{math}
P \left[ \left\{ s_i \right\} \right] = 
\frac{ \prod_i{ exp( h_i s_i ) } }
{\sum_i{exp(-E \left[ s_i \right] ) / T } }
\end{math}
 
We let a ``partition function'' be given by:\\
\begin{math}
Z = \sum_i{exp(-E \left[ s_i \right]) / T }
\end{math}
 
then we can find the minimum energy by computing the partial derivative of the natural logarithm of $Z$ with respect to the inverse of temperature $\beta$. \\
 
\begin{eqnarray}
\ln(Z) & = & \ln \left( \prod_i{exp(\beta h_i) + exp(- \beta h_i)} \right)  \nonumber \\
& & = \sum_i ln { \left[ exp(\beta h_i) + exp(- \beta h_i) \right]  } 
\end{eqnarray} 
 
\begin{figure}
\centering
\includegraphics[height=\columnwidth,angle=270]{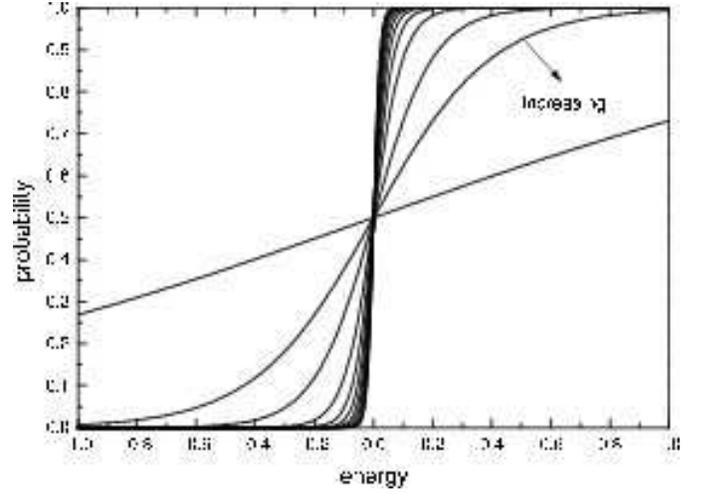}
\caption{Magnetic domain ``transfer function''}
\label{fig:Magnetic_domain_transfer_function}
\end{figure}

From this we can write the minimum energy as

\begin{eqnarray}
u & = &  \frac{ \delta \ln(Z) }{ \delta \beta} \nonumber \\
& & = \sum{\frac{h_i(exp( \beta h_i) - exp(- \beta h_i))}{(exp(\beta h_i) + exp(- \beta h_i))}}  \nonumber \\
& & = \sum_i{ h_i \tanh(\beta h_i)}
\end{eqnarray}
\\
Figure \ref{fig:Magnetic_domain_transfer_function} is a plot of this function.

This is a very interesting representation for the magnetic field, because it parallels computational systems known as neural networks (cf. \cite{Hopfield:1982} and \cite{Bienenstock:1986}). We can write the magnetization $m$ as
          
\begin{equation}
m = \left\langle s_i \right\rangle = \tanh(\beta h_i)
\end{equation}
          
This equation says the hyperbolic tangent of spin energy divided by the temperature gives the magnetization. For neural networks the output of a neuron is given by the hyperbolic tangent of the sum of the inputs to that neuron. As is well known the hyperbolic tangent function is graphically represented by a function exactly like that shown in Figure \ref{fig:Magnetic_domain_transfer_function}.

We now expand the model to allow the spins to interact with their neighbors. The interaction energy, $J$, between spins may come about through several mechanisms, but the end result will be that the spins will take on either a direction $J>0$ or $J<0$. In a similar way to the derivation of the energy relation of Eq \ref{ising1} we can write an energy relation including the spins and interaction energy, $J$ and get the following equation:

\begin{eqnarray}
E \left[ \left\{ s_{ij} \right\} \right]  = &  - \sum_{ij}{h_{ij}s_{ij}} - & \nonumber \\
 & J \sum_{ij}{( s_{ij}s_{i+1j} + s_{ij}s_{ij+1} )} &
\end{eqnarray}

This equation includes more dimensions because now we are including the neighbor interactions. In the previous equation we did not include neighbors so the current equation appears to be more complex.

%
%
%
%
%
%
%
%
%

Briefly, here is how we can use Ising spins for computation. We will view a magnetic quantum dot, Ising array as a dynamical system that is carrying out some specific computational task. For example, by associating some input state with a reproducible output state we can build an associative memory.

We propose using an array of magnetic quantum dots that have locally programmable regions of the dot array. Various regions can be programmed with magnetic probes to induce specific magnetic spins in the dots. These dots can then interact with their nearest neighbor as a complex dynamical system. By using an array of magnetic probes on the edges as input and output channels, we can exploit the entire landscape of magnetic quantum dots for analog computation. Furthermore, the system is fault tolerant. If segments of the array are damaged, or nonfunctional, the array is dynamic and can adapt to these defects.

\begin{figure}
\centering
\includegraphics[height=\columnwidth,angle=270]{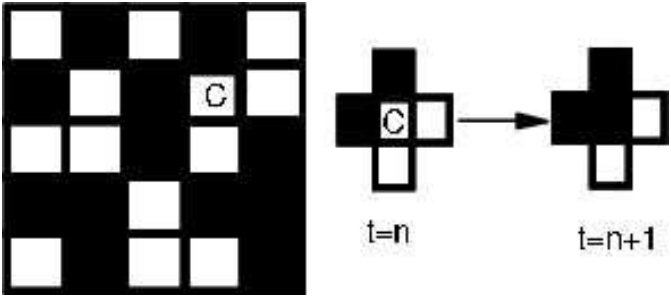}
\caption{Spin interaction map}
\label{fig:Spin_interaction_map}
\end{figure}

\subsection{Simulation of Magnetic Dot Computing}

A cellular automaton is an array of small computational elements called finite state machines. Each cell in the array is capable of receiving input from its nearest neighbors only. At each time increment (time is discrete), the input from the neighbors determines the new state for the cell. For example, consider the small array in Figure \ref{fig:Spin_interaction_map} and the cell marked C in that array. At time $t=n$ we see that cell C is white and is surrounded by two black cells. At time $t=n+1$ we see the center cell is now changed to a black state. This change is due to dynamic interactions between the four neighbors and the center cell. Every cell in the entire array is updated and the updating can be synchronous or asynchronous. If synchronous, than the next state for each cell is computed before any changes take place in the array. At that time all the cells undergo the computed state change. If the dynamics are asynchronous then the computations are done for a cell selected at random and the cell undergoes the required change. This in essence, is a cellular automaton. 

Both electronic quantum dot arrays (\cite{Porod:1999}) and a magnetic quantum dot arrays are analogues to cellular automata.  In each case, the computations carried out at each "dot" is done by nearest neighbor interactions. The electronic quantum dot array can be used for computational purposes by electronic interaction between nearest neighbor quantum wells. The magnetic quantum dot array can be used for computational purposes by nearest neighbor interaction of the magnetic domains. All of these systems have been modeled using cellular automata. 

As pointed out in \cite{Landauer:1995} and \cite{Roychowdhury:1997}, the primary problem associated with using cellular automata or dot arrays for computation is the fact that the flow of information is bidirectional. There is no isolation between the inputs and the outputs. Here we present an approach suggested in \cite{Roychowdhury:1997}, but we include enough detail to make the suggested methods practical, and we present simulations to demonstrate feasibility. Though we expect the method will also work with regular electronic quantum dots, cellular automata, and molecular oscillator-based computing. The key to utilizing these systems for computational purposes, is to recognize that the dynamics of the system are sensitive to initial conditions. This is equivalent to setting up the input data for a computation. The evolution of the dynamics of the computation will be adjusted by controlled perturbations of the quantum dots. These perturbations are equivalent to programming the system. And the final spatial configuration is the computed answer to some mathematical mapping.

So summarizing the dynamics of the Ising computer we can rewrite the relation for energy of the magnetic interaction as

\begin{equation}
\label{magdot1}
E_{ij} = -(B_{ij} + J_{ij} s_i s_j)
\end{equation}

That is, the energy at each site is given by the product of the neighboring spins and the interaction energy $J$. Here we have also included an external magnetic field $B$. The probability of a given spin state at some site is given by
    		
\begin{equation}
\label{magdot2}
P(s_i) = \frac{1}
{1 + exp \left[ - \left( \sum_j{J_{ij} s_i s_j / T }\right) \right] } 
\end{equation}
    		
This has the same functional form as the plot shown in Figure \ref{fig:Magnetic_domain_transfer_function}. The figure is a plot of the probability at several temperatures. At very low temperatures, the probability will tend to take on either zero or one. At warmer temperatures the probability will tend more-and-more toward $P=0.5$.  These two equations (Eq [\ref{magdot1},\ref{magdot2}]) are the only ones needed for a simulation of the Ising Computer.

\begin{figure*}
\centering
\includegraphics[height=5in,angle=270]{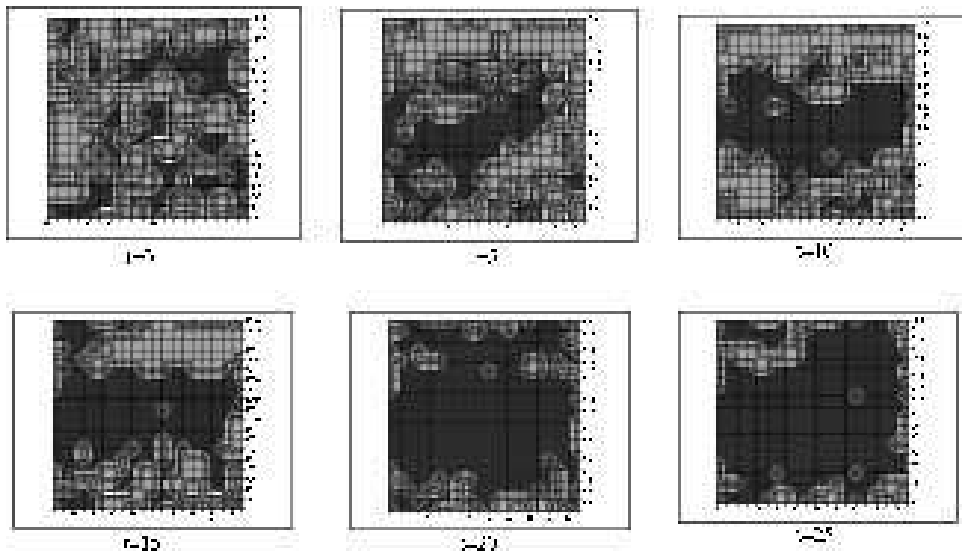}
\caption{Time dynamics of the magnetic quantum spins.}
\label{fig:Time_dynamics_of_the_magnetic_quantum spins}
\end{figure*}

\begin{figure*}
\centering
\includegraphics[height=5in,angle=270]{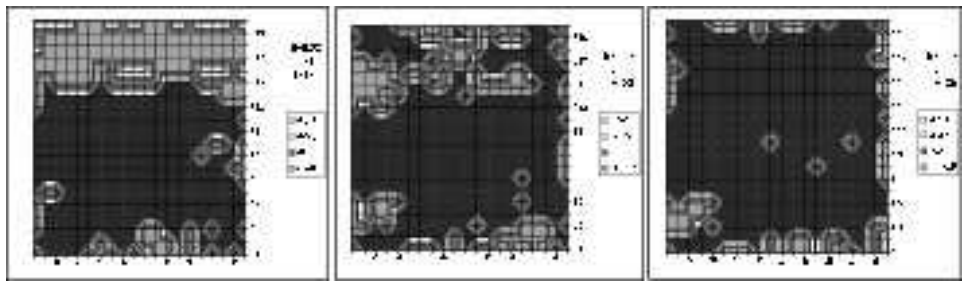}
\caption{Temperature study of magnetic spin quantum dot computer.}
\label{fig:Temperature_study_of_magnetic_spin}
\end{figure*}

Typically in running simulations of Ising systems one would use periodic boundary conditions. Wrapping the edges will accomplish this, so that the top and bottom are connected, and the left and right edges are connected. Since we want to use the system for computation we will use the edges for input and output, so we will not use the periodic boundary condition. Figure \ref{fig:Time_dynamics_of_the_magnetic_quantum spins} shows the spins for a $20 X 20$ system at various snap-shots in time.  We see from Figure \ref{fig:Time_dynamics_of_the_magnetic_quantum spins} that as the dynamics progresses in time the spins become more-and-more locally correlated. The effects of temperature are shown in Figure \ref{fig:Temperature_study_of_magnetic_spin}. As indicated in Equation \ref{magdot2} and Figure \ref{fig:Temperature_study_of_magnetic_spin}, the temperature affects the dynamics.

To demonstrate the programmability of the system we used three edges as the input and one edge as the output. We used a genetic algorithm (cf. \cite{Goldberg89}) with three, 2-dimensional chromosomes per member in a population of Ising systems. The three chromosomes represent the initial configuration, $I$[x][y], the coupling magnetization between sites, $J$[x][y], and an added magnetic field, $B$[x][y], to increase the dynamic programming ability of the system. The relevant dynamics is given by the probability 

\begin{equation}
\label{magdot3}
P_i = \frac{1}
{1 + exp\left[ - \left( B_i + \sum_{i=0}^{i=4}{J_{ij} s_i s_j } \right)/ T \right]}
\end{equation}

This form of the equation hardly differs much from Equation \ref{magdot2}. The only difference is the inclusion of the external magnetic field $B$.

\begin{figure}
\centering
\includegraphics[height=\columnwidth,angle=270]{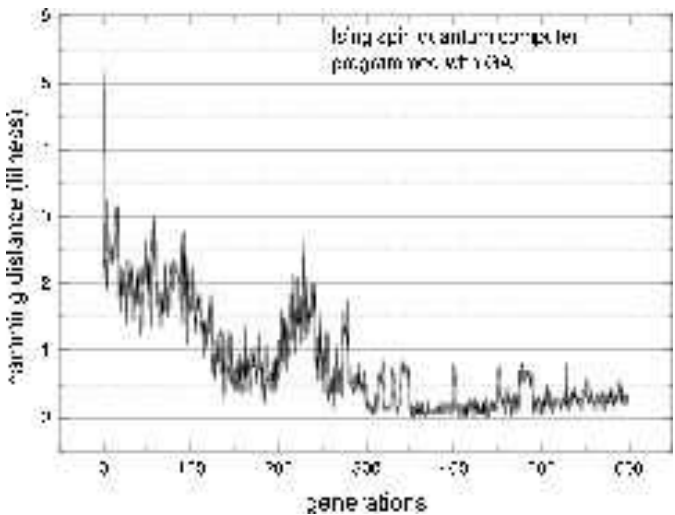}
\caption{Learning curve for pattern matching problem using magnetic quantum dot computer.}
\label{fig:Learning_curve_for_quantum_dot_computer}
\end{figure}

Figure \ref{fig:Learning_curve_for_quantum_dot_computer} shows the results from a simulation. The system works as follows. The genetic algorithm manipulates the elements of the three matrices, $I, J, B$ such that a particular input pattern can be mapped to a particular output pattern with a low Hamming distance between the actual output and the target-output pattern. In a sense, the system acts like a feedforward neural network. The information on the input edges is dynamically interacting with their nearest neighbors. These neighbors in turn are interacting deeper into the array. So the information from the input edges is essentially fed into the array. The array relaxes to some configuration with the output edge giving the final computation.

In the initial experiments all three matrices $(I, J, B)$ were initialized by the genetic algorithm and the system was allowed to settle to the answer, some pre-selected pattern on the output. It should be possible to also use more complex arrays of $J(t)$ or $B(t)$ and to let these represent a dynamic program. For example, with an array size of $20 X 20$ one would have an active dynamic array of $19 X 19$, assuming all edges are used for input/output, and then the arrays J and B would be configured as follows. $J[x][y][t]$ and $B[x][y][t]$ where each x,y array also has a time index from $t=1$ to $t=19$. This is easily computed by the fact that the signal can only travel one spatial site per time step (called the speed of light). Of course it should be possible to use any $t$ index depending on the problem. For example, some problems may require only $t-n$ updates of the $B$ and $J$ matrices, while others may require more. It could also be possible to run the dynamical system until $n$ time units after the $B$, $J$ updates to the program. In this case the free running time after set-up, would be part of the programming of the system. 

\subsection{Prototype System}

In the above simulation we made several assumptions, primarily concerning size of the magnetic dots and how the dots would interact with each other. First, we assumed that the dots would be larger in diameter than in height. This important assumption will result in the dots taking only one of two states (spin up and spin down). Shinjo et al. first observed this experimentally (2000)\cite{Shinjo:2000}. Their observations indicate that dots on the order of microns in size (1-2 microns) will still exhibit only two states. Dots this large suggest that we can make a prototype easily.

We know that the magnetic quantum dots would not really be able to implement quantum computation. The magnetic dots are too large to be called quantum dots. Classical electrostatics will dictate the overall dynamics. Cowburn et al. describe room temperature quantum cellular automata made with magnetic quantum dots\cite{Cowburn:2000}. When a cluster of dots takes on the same spin state it acts essentially as a single giant classical spin system. Their quantum dots were about 110 nm in diameter and 10 nm thick.

As demonstrated in \cite{Shinjo:2000} and \cite{Cowburn:2000}, small magnetic fields are needed (e.g. 300 Oe) for programming these dots. Both sets of investigators use small magnetic ``pins'' to inject fields into specific locations. Shinjo et al. uses a magnetic probe to monitor the spin state and Cowburn and Welland use reflected light to access the longitudinal Kerr effect to observe the spin state. 

\begin{figure}
\centering
\includegraphics[height=\columnwidth,angle=270]{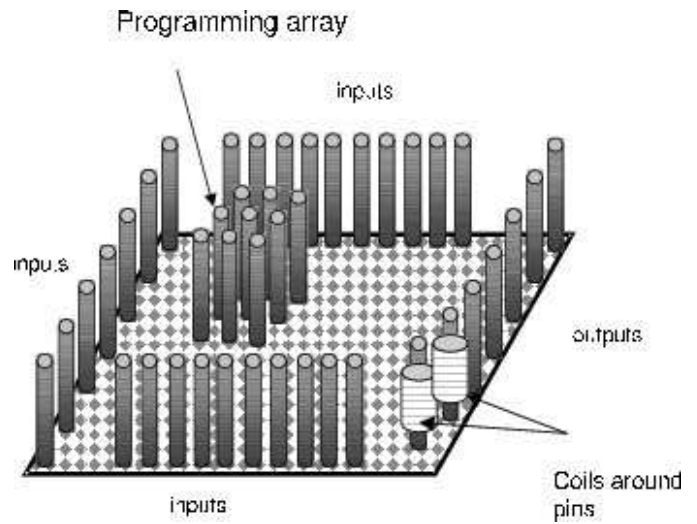}
\caption{Schematic of a prototype for magnetic quantum dot computing.}
\label{fig:prototype_for_magnetic_quantum_dot_computing}
\end{figure}

Figure \ref{fig:prototype_for_magnetic_quantum_dot_computing} shows a schematic of a prototype system that could easily be constructed. Some of the edges would be inputs and another edge would be the outputs. The edge probes are held fixed in position and the inputs are held at a positive or negative voltage so as to maintain the input state to the array. The output probes continually monitor the output dots. In the center of the array is a small cluster of magnetic probes that raster back-and-forth to program the initial state. 

Using this technique to raster a cluster of probes, in which the probe size is potentially larger than the magnetic quantum dots, we can still perform useful computations. In our simulation we assumed that we could not necessarily access each dot, but rather a small cluster of dots with each magnetic probe. In other words, the probe size was larger than the dot size.

In \cite{Cowburn:2000} and \cite{Prinz:1998} a technique is described that uses RAM cells to program the individual dots. They calculate that a microprocessor based on magnetic quantum dots would draw only one Watt of power. Of course, electronic control of the dots is also possible, for example spin transistors\cite{Prinz:1998} and \cite{Gorelik:2003}.

\section{Programmable Fermi Surfaces}

Consider a programmable resistor array. It can be used to perform the mathematical operation of vector-matrix multiplication. This operation is at the heart of target recognition, pattern recognition, data compression, process control, associative memory, content addressable memory, and many other computations. In fact, vector-matrix multiplication is the primary mathematical operation that digital signal processing (DSP) chips undertake.

\begin{figure}
\centering
\includegraphics[height=\columnwidth,angle=270]{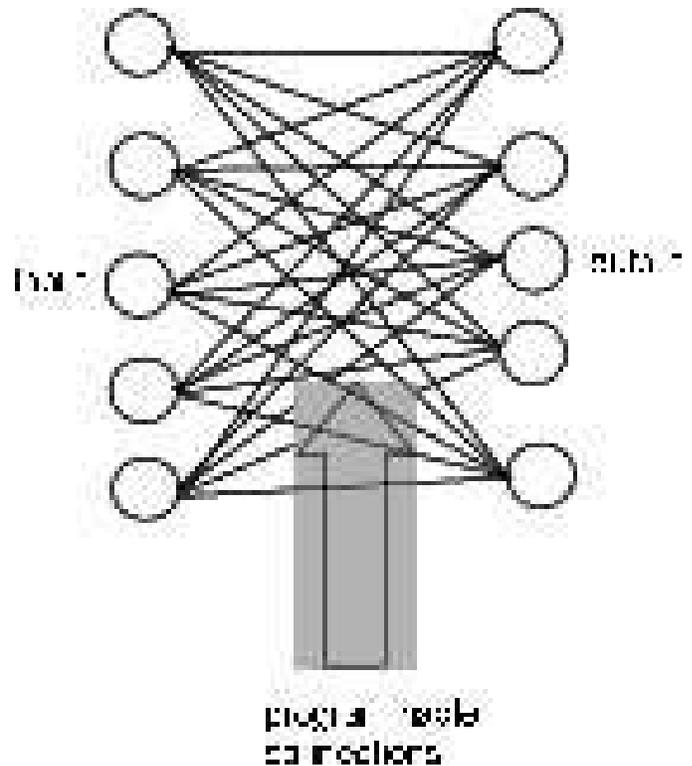}
\caption{Schematic of a programmable resistor array.}
\label{fig:Schematic_programmable_resistor_array}
\end{figure}

To exploit a programmable resistor array for associative memory (a type of neural network), for example, we recall that an associative memory is a mapping between two different vectors. When input nodes in the circuit are activated the circuit will automatically perform a vector-matrix multiplication and the output nodes will present the nearest mapping relations to the input vector. The matrix of connections can be pre-computed so that the system will respond to several (or many, depending on the system size) different input stimulates.  

The same system can be used for pattern recognition. The target pattern will be coded as a vector. Then compute the cross product of that vector and its transpose. This will generate a matrix. This matrix is the resistance values for the circuit shown in Figure \ref{fig:Schematic_programmable_resistor_array} - the programmable resistor array. When vectors similar to the expected target are presented to the inputs the circuit will compute a vector matrix multiplication and complete the recognition of the target vector (or image).  All these computations are well known and fall into the category of neural networks (\cite{Hassoun:1995}; \cite{Rumelhart:1986}). These types of computations are also exactly the type that can be done with a programmable crystal lattice.

\begin{figure}
\centering
\includegraphics[height=\columnwidth,angle=270]{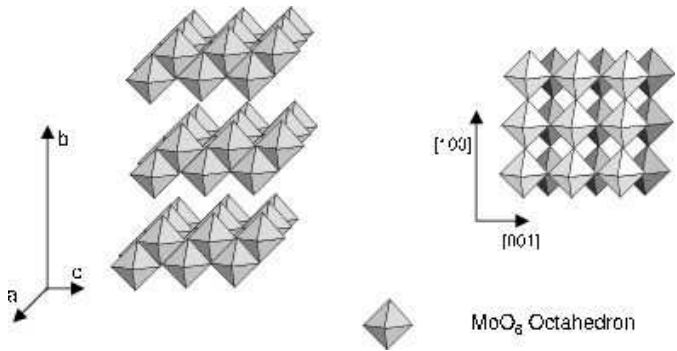}
\caption{Crystal structure of $Mo O_6$ (side view, top view)}
\label{fig:Crystal_structure_of_MoO6}
\end{figure}

\begin{figure}
\centering
\includegraphics[height=\columnwidth,angle=270]{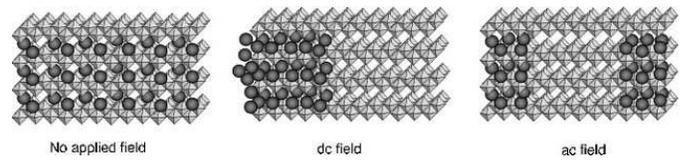}
\caption{Effects of no field, dc field and modulated ac field on the ions in the lattice.}
\label{fig:Effects_of_fields_ions_in_the_lattice}
\end{figure}

Certain materials crystallize into stacks of 2-dimensional layers (Figure \ref{fig:Crystal_structure_of_MoO6}). One example of this class of materials is molybdenum trioxide, an insulating material. It is possible to electrochemically reduce the molybdenum while it is in the crystal lattice and intercalate ionic species, such as $Li^+$, $Na^+$, $K^+$ or $Cs^+$ into the lattice (Figure \ref{fig:Effects_of_fields_ions_in_the_lattice}). The material will then become a semiconductor and fast-ion conductor. 

$K_{0.3} Mo O_3$ is an excellent example of a fast-ion conductor. Pure $Mo O_3$ is white to yellow in color. The potassium-doped crystal is dark red to purple in color; this darker color indicates that the electronic conductivity has increased. The doped crystal is a mixed ionic and electronic conductor.  It is electronically a semiconductor and a potassium ion conductor. The potassium ions disrupt the electron density (the Fermi surface) of the molybdenum atoms. Of course the electron density is an image of the charge density.

In an ac field the ions will drift back-and-forth at the frequency of the applied field (Figure \ref{fig:Effects_of_fields_ions_in_the_lattice}). So in a frequency and amplitude modulated ac field, the ions can be moved to within a few nanometers of each other to form clusters and then ``pinned'' into place with a higher frequency field. Naturally these clusters of pinned ions would prefer to move away from each other and settle into some equilibrium state commensurate with the underlaying molybdenum oxide lattice. These metastable clusters of ions disturb the underlaying Fermi surface of the molybdenum ions, and generate waves of disturbance with a correlation length on the order of 100s of microns. These are known as charge density waves and they can be measure by their effects on ac and dc fields. Basically, by moving these ions in an ac field and pinning them we are in effect programming Fermi surfaces.

\subsection{Simulations}

In the absence of an electric field the ions will repel each other to a maximum distance and form a superlattice that may be commensurate with the under-laying primary lattice. In the presence of electric fields the ions will move and the superlattice will now become incommensurate. The lattice vibrations in this incommensurate region of the crystal will send out quasi-periodic or even chaotic waves (\cite{Wilson:1975}; \cite{Gruner:1994}). It is believed that each of these ionic centers forms a phase vortex ring (\cite{Gill:1986}). These vortex rings will expand until either a defect or another vortex ring is encountered. They can have a diameter of up to 100s of microns depending on the number-density of vortex rings (\cite{Fukuyama:1978}; \cite{Gruner:1994}). The rings behave like quasi-periodic oscillators, known as sine-circle maps (\cite{Azbel:1984}; \cite{Zettl:1986}; \cite{Sherwin:1988}; \cite{Inui:1988}; \cite{Gruner:1994}). Obviously, a molecular scale array of them is a type of cellular automata known as a coupled map lattice. The number-density, which can be controlled by the stoichiometry or by the electric field modulation, will essentially dictate the size of the rings. We can exploit these phenomena for computation. 

We can ``program'' the vortex ring array by moving the ions (the vortex ring generating entities) with modulated ac fields. The ions can then be ``pinned'' into place with high frequency fields - about 10 MHz (\cite{Cava:1984,Cava:1985}. The rings can be ``connected'' with external fields (\cite{Csiba:1989}; \cite{Gruner:1994}; \cite{Karttunen:1999}; \cite{Gill:1986}; \cite{Hall:1988a,Hall:1988b}). At a specific threshold the rings will produce a sliding charge density wave. The actual ``programming'' can be done with a genetic algorithm searching for a ``circuit'' that computes a predefined mapping relation (exactly as was done with a FPGA search for oscillators). The best analogy is to think of this as nanoscale programmable resistor array - or programmable crystal lattices.

Several interesting models of this phenomenon have been discussed in the literature. The simplest model is based on Hamiltonian dynamics of particles trapped in wells (\cite{Myers:1993}). The model starts by assuming a Hamiltonian lattice to describe the effects of an applied electric field. The charge density wave elasticity and the pinning disorder are given by

\begin{eqnarray}
H  = & - F \sum_i{x_i} + \nonumber \\
& \frac{1}{2} J \sum_{ \left\langle ij \right\rangle} {(x_i - x_j)^2} - & \nonumber \\
& \frac{V}{2\pi} \sum_i{\cos \left[ 2\pi (x_i - \beta_i) \right]  }
\end{eqnarray}
 
where $\beta_i$  is a random number between 0 and 1 selected for each site, $i$, but not at each time iteration. $F$ is the applied field; $V$ is the pinning potential, $J $is the coupling strength (like a spring constant); and $x_i$  is the phase at site $i$. The dynamics evolves according to

\begin{math}
{\mathop x\limits^.} _i  = F - V\sin \left[ {2\pi (x_i  - \beta _i )} \right]
\end{math}

\begin{figure}
\centering
\includegraphics[height=\columnwidth,angle=270]{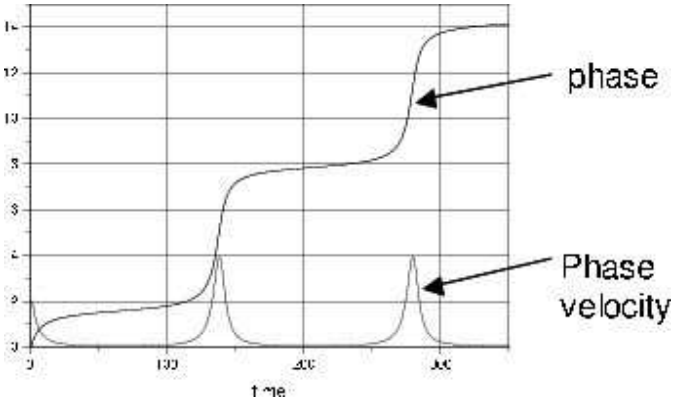}
\caption{Myers and Sethna model}
\label{fig:Myers_and_Sethna_model}
\end{figure}

The equation has solutions at zero when $F \leq V$ . The solutions are shown graphically in Figure \ref{fig:Myers_and_Sethna_model}. The figure shows the phase and the phase velocity. The dynamics suggests a stick-slip behavior. The charge density wave gets stuck at some pinning center, and when the field increases, it will rapidly unstuck or slip and then get stuck at another point.

The Myers and Sethna model is very interesting and has some similarity to the models of \cite{Azbel:1984}; \cite{Zettl:1986}; \cite{Sherwin:1988}; \cite{Inui:1988}; \cite{Gruner:1994} who discuss sine-circle maps. Specifically Azbel and Bak suggest a model of the following type. 
  	
\begin{math}
x_{n+1} = x_n + \Omega + ( \kappa / 2 \pi ) \sin(2 \pi x_n)
\end{math}
  	
Azbel and Bak did not assemble array models of this equation where each cell in the array was a separate vortex interacting with their neighbors. In order to do that we will draw on the work of Kanek, who describes a simplified version of this equation in a cellular array known as a coupled map lattice (like a real-number cellular automata)\cite{Kanek:1991}. We will make further assumptions based on the work of \cite{Sinha:1998} who worked with coupled map lattices, and had the objective of demonstrating computation. In brief, we will combine the equation of Kaneko with the algorithmic modifications (to be described subsequently) of Sinha and Ditto and model computation with these sine-circle maps.

The Kaneko modification of the sine-circle map (hereinafter called circle map) is given by

\begin{math}
x_{n+1} = x_n + ( \kappa / 2 \pi ) \sin(2 \pi x_n)
\end{math}
 
We further simplified this equation as follows:

\begin{math}
x_{n+1} = x_n + K \sin(Nx_n)
\end{math}
 
\begin{figure}
\centering
\includegraphics[height=\columnwidth,angle=270]{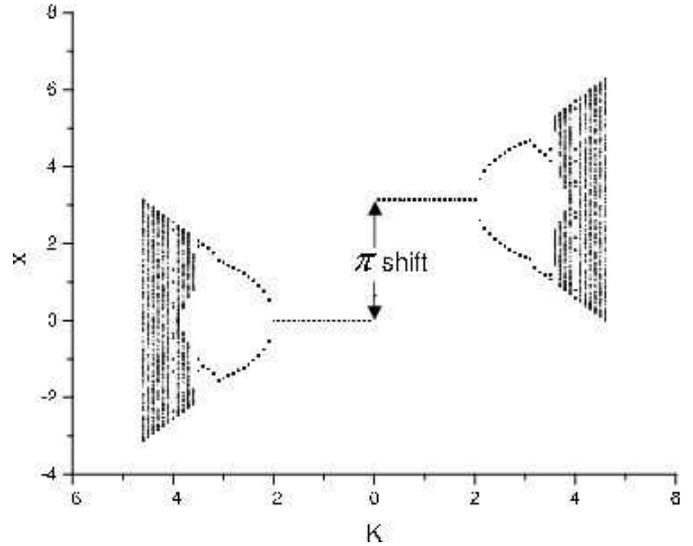}
\caption{Bifurcation diagram for circle map.}
\label{fig:Bifurcation_diagram_for_circle_map}
\end{figure}

The ``bifurcation diagram'' shown in Figure \ref{fig:Bifurcation_diagram_for_circle_map} outlines the dynamics of this equation. Here we let $N=1$. From the bifurcation diagram it is clear a whole range of dynamics is possible. Zero is the attracting point from $K = -2$ to $K = 0$. A phase shift of  occurs at $K = 0$ and remain at that level untill $K = 2$.

\begin{figure*}
\centering
\includegraphics[height=5in,angle=270]{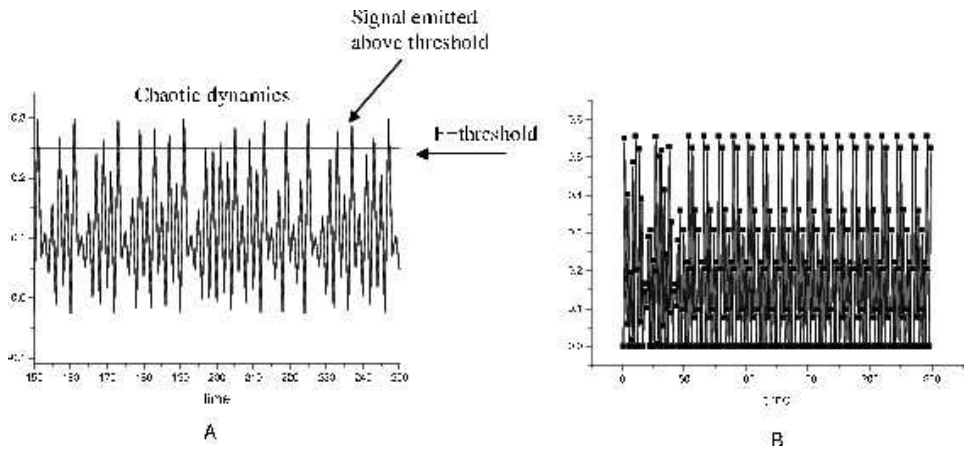}
\caption{Sinha and Ditto model adapted for charge density wave computation (see text for discussion)}
\label{fig:Sinha_and_Ditto model}
\end{figure*}

If each vortex site in the crystal lattice behaves like this model we can simulate an array of them and couple them together. This is called a coupled map lattice. To couple them together we will use an external threshold, $T_n$ at each site. Connecting a linear array of these to a DC threshold we obtain the results shown in Figure \ref{fig:Sinha_and_Ditto model} (see \cite{Sinha:1998} for algorithm details).

In this simulation the cells were operating with identical conditions and were in the chaotic regime. Figure \ref{fig:Sinha_and_Ditto model}a shows the chaotic signal and a threshold line. Whatever signal is above threshold will carry over into the next cell in the linear array (in a feed forward manor) and will add to the equation.  Doing this, and monitoring the pulses at the last node in a chain of 100 cells, we find that periodic pulses develop after a short time interval. The figure on the right (Figure \ref{fig:Sinha_and_Ditto model}b) shows the periodicity. This is exactly one of the defining behaviors for charge density wave materials - periodic pulses are observed in a DC field (\cite{Cava:1985}; \cite{Cava:1984}).

\begin{figure}
\centering
\includegraphics[height=\columnwidth,angle=270]{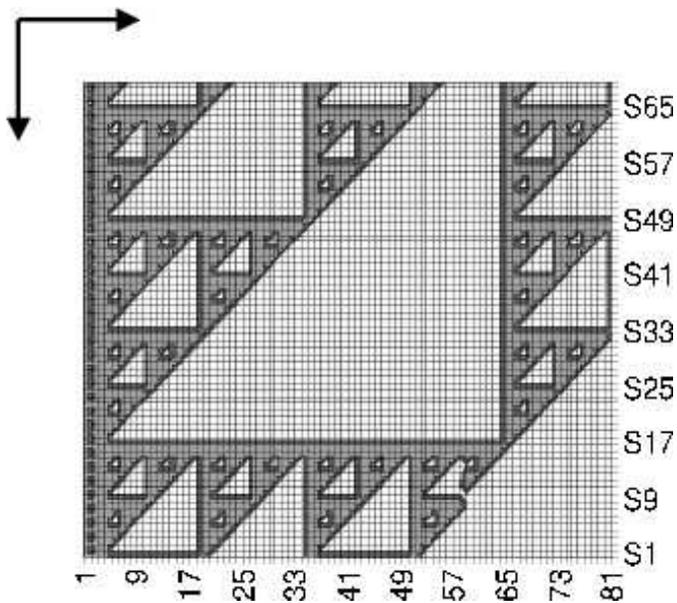}
\caption{Two-dimensional lattice simulation of charge density waves using circle-map equation and threshold dynamics of Sinha and Ditto (1998)\cite{Sinha:1998} to represent an applied dc field}
\label{fig:Two-dimensional_lattice_simulation}
\end{figure}

With a two-dimensional array and two equal but perpendicular dc fields we expect to see traveling pulses above threshold. Figure \ref{fig:Two-dimensional_lattice_simulation} shows this phenomenon. The figure shows a two-dimensional plot of the above threshold values coming out of the oscillators. Periodicity in two-dimensions is clearly seen. It represents the ``stick-slip'' phenomena and gives rise to the sliding charge density wave. 

\begin{figure*}
\centering
\includegraphics[height=\columnwidth,angle=270]{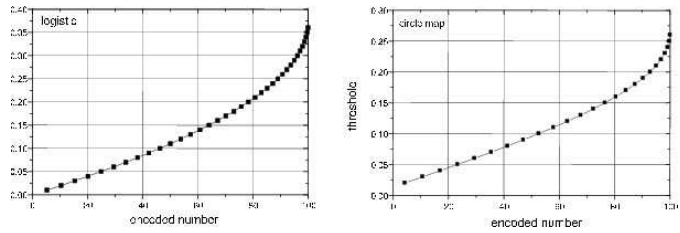}
\caption{Logistic- and circle-map adding machines from linear arrays (see Sinaha and Ditto, 1998)}
\label{fig:Logistic_and_circle-map_adding_machines}
\end{figure*}

Sinha \cite{Sinha:1998,Sinha:1999,Sinha:2002a,Sinha:2002b} demonstrates, in simulation, that chains of logistic functions can be used for Boolean logic and numerical calculations. For example, it is possible to build an adding machine by setting individual thresholds at each machine, and that the emitted signal at the end of the chain is equal to the addition of the thresholds. The mapping relation encoded the numbers and threshold is given in Figure \ref{fig:Logistic_and_circle-map_adding_machines} for both chains of logistic and circle map functions. Further, as to be expected arrays of these oscillators can be configured for constructive and destructive interference of pulses to produce Boolean logic functions.

\subsection{Prototype System}

\begin{figure*}
\centering
\includegraphics[width=5in]{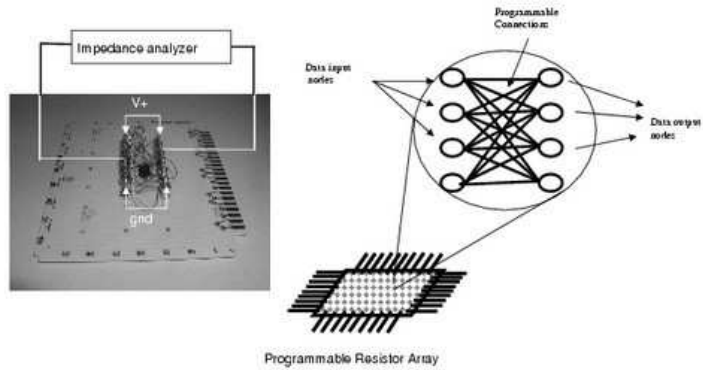}
\caption{Simple programmable resistor network from charge density wave material}
\label{fig:programmable_resistor_network}
\end{figure*}

Single crystals and epitaxy films will exhibit the phenomenon of charge density waves that can be exploited for analog computation by pinning the ions in specific locations. These materials were first investigated for solid-state batteries as an electrolyte, because they are ionic conductors. Obviously prototype batteries were built from compressed powders of $K_{0.3}MoO_3$. In order for these materials to work as batteries the ions must pass from grain-boundary to grain-boundary. This implies that we do not even need single crystals (though no doubt epitaxy layers would be ideal) for preliminary experiments. Though in this situation we would not expect perfect control of the patterns of ions. Yet even with this we can exploit this phenomenon for computational purposes. The system, would act like a feedforward neural network \cite{Rietman98} that can be programmed with a genetic algorithm. The program would be adjusting, essentially, tiny resistors. Figure \ref{fig:programmable_resistor_network} shows a schematic diagram of our system and a prototype to test the basic ideas.

\begin{figure}
\centering
\includegraphics[angle=270,width=\columnwidth]{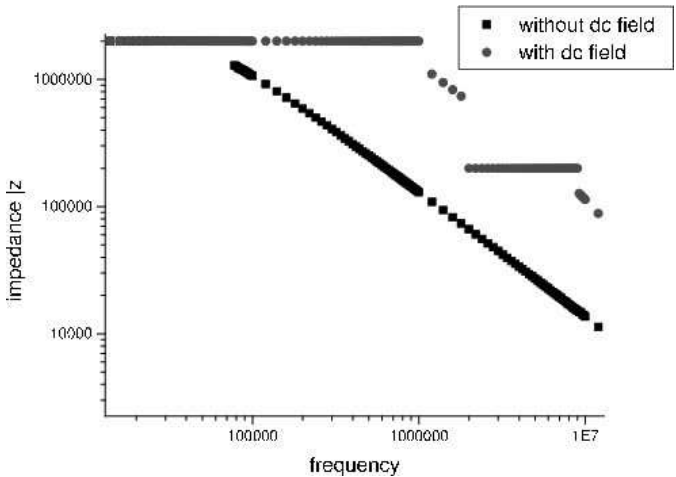}
\caption{Results from prototype of the CDW device}
\label{fig:Results_from_ prototype_of_the_CDW_device}
\end{figure}

Of course, if the material is polycrystal this will add uncontrolled defects to our programmable surface. These crystal boundaries will result in faults in the computational network. However, as long as we take these faults into account with our design algorithms (a genetic algorithm) it should have little or no impact. Figure \ref{fig:programmable_resistor_network} shows a prototype system we built to test the basic idea of programmable resistor arrays. A film of $K_{0.3}MoO_3$ was made with 5 wt\% polyethylene oxide as a binder and deposited on a glass slide with previously deposited gold electrodes with wires attached. The wires were connected to a simple screw terminal for strain relief. The two end terminals were connected together and then connected to a dc field (20 volts). Terminals from opposite sides were then connected to a 4192A impedance analyzer and the ac impedance was measured as a function of frequency. Figure \ref{fig:Results_from_ prototype_of_the_CDW_device} shows these, very preliminary, results.  

The results indicate that there was more than an order of magnitude change in the resistance with the dc field on versus the dc field off. No doubt if we used epitaxy films (\cite{Mantel:1997}; \cite{Zant:1997}) we could have better control of the ions.

\section{Acoustic Modulation of Media for Optical Computing}

It is possible to use acoustic pulses, in particular standing waves, for modulating the density of solids, liquids, and gases. In this section we review  work by Higginson et al\cite{Higginson:2004a, Higginson:2004b}, that describe the use of acoustic standing waves to modulate the density of media. 

The interaction between sound and optics was first described in \cite{Brillouin:1922}, and later elaborated on in \cite{Deby:1932}. Normally acousto-optic effects arise form traveling acoustic waves producing a periodic modulation in the refractive index of the medium. This effect is well known for surface acoustic wave devices and produces a grating. The traveling acoustic wave has a much lower velocity relative to the speed of sound so it effectively acts as a grating. The effect has been exploited for an acousto-optic television display \cite{Korpel:1966}. This television application suggests other technologies such as optical switching and mask-less lithography. These systems exploit a traveling wave. Contrast this with the work in \cite{Higginson:2004a, Higginson:2004b} who have been making use of an acoustic standing wave.

\begin{figure}
\centering
\includegraphics[height=\columnwidth,angle=270]{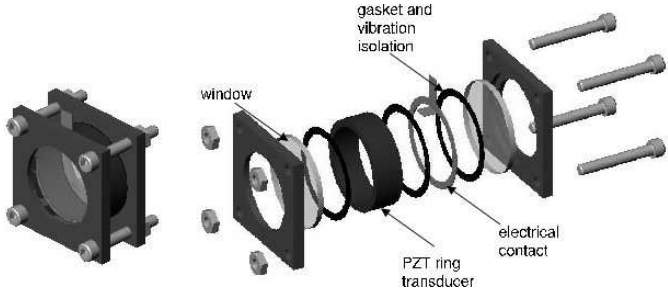}
\caption{Acusto-optic device for lensing, described by \cite{Higginson:2004a, Higginson:2004b}}
\label{fig:Acusto-optic_device_for_lensing}
\end{figure}

In order to exploit an acoustic standing wave for modulating light they started with cylindrical device containing fluid (see Figure \ref{fig:Acusto-optic_device_for_lensing}). One starts the analysis by considering the time-independent component of a standing wave, which can be derived from the Tait equation for liquids \cite{Hamilton:1998}). After expanding by a second order Taylor series we get

\begin{math}
\rho - \rho_0 = \frac{1}{c^2_0}(P-P_0)  + \frac{1 - \gamma}{2 \rho_0 c^4_0}(P-P_0)^2 + \ldots
\end{math}
 
 where $P$ and $\rho$  are the total pressure and density,$P_0, \rho_0, c_0$  are the ambient pressure, density, and sound speed. $\gamma$ is an empirical constant.

Since the pressure is adiabatic, and the flow is irrotational, in the cylindrical cavity, expanding the total pressure in a Taylor series and grouping the terms can estimate the pressure terms. Time-averaging eliminates the first-order terms and gives the time-invariant component of the density in the liquid.

\begin{math}
\rho_2 = \left\langle \rho - \rho_0 \right\rangle =
\frac{2 - \gamma}{2 \rho_0 c^4_0} \left\langle \rho^2 \right\rangle - 
\frac{\rho_0}{2c^2_0} \left\langle u \bullet u \right\rangle
\end{math}
 
where $p$ and $u$ are first-order acoustic pressure and velocity, obtained by solving the linearized wave equation, and the angular brackets indicate time average. Since the geometry is circular the acoustic standing wave can be described by
 
\begin{math}
p(r,t) = AJ_0(kr)\sin(\alpha t)
\end{math} 

\begin{figure}
\centering
\includegraphics[height=\columnwidth,angle=270]{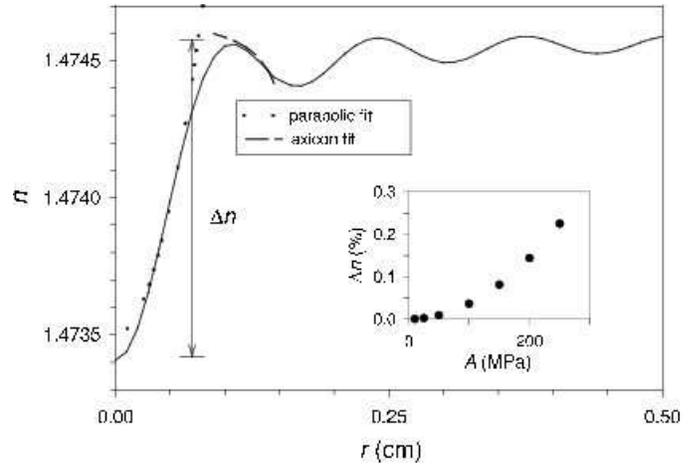}
\caption{Calculated refractive index profile in glycerin for A=200 MPa and $\Omega \ 2 \pi = 700kHz$ . The inset is the refractive index disturbance as a function of acoustic pressure.}
\label{fig:Calculated_refractive_index_profile_in_glycerin}
\end{figure}

So we get
\begin{math}
\rho_2(r) = \frac{A}{4 \rho_0 c^2_0}
\left[
(2 - \gamma) J_0(kr) - J^2_1(kr)
\right]
\end{math}

where $A$ is the amplitude of the sound pressure, $\alpha$ is the angular frequency, $k$ is the dispersion-free wave number $\omega / c_0$, $J_i$  are Bessel functions of the first kind, and $r$ is the radial coordinate.  The density is related to the refractive index $n$, by Lorenz-Lorenz equation. We can write the time averaged version of this as
 
\begin{eqnarray*} 
\left\langle n \right\rangle = 
n_0 + \alpha \left\lfloor (2 - \gamma) J_0 (kr) - J^2_1(kr) \right\rfloor
\\
\alpha = \frac{A^2 Q(n^2_0 +2)^2}{24 \rho - \rho_0 c^4_0 n_0}
\end{eqnarray*}
 
where $Q$ is the molar refractivity of the material, and  $n$ is the refractive index at $\rho_0$.
The calculated time-averaged index profile of a glycerin-filled circular cavity is shown in Figure \ref{fig:Calculated_refractive_index_profile_in_glycerin}. Here we have used the values
\\
\\
\begin{math}
\rho_0 = 1260 kg \ m^3  \\
c_0 = 1904 m \ s \\
n_0 = 1.4746 \\
Q = 2.23X10^{-4} m^3 \ kg \\
\gamma = 10.0 \\
\Omega \ 2 \pi = 700 kHz \\
A = 150 MPa 
\end{math}

\begin{figure}
\centering
\includegraphics[height=\columnwidth,angle=270]{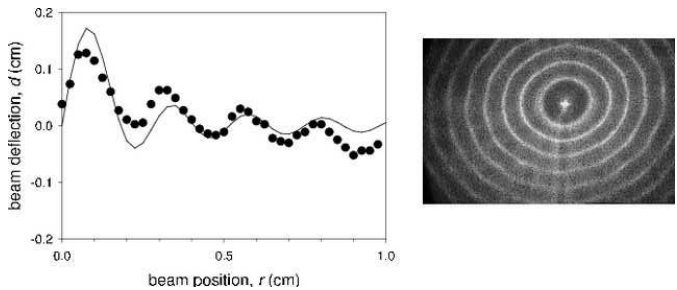}
\caption{Predicted and actual beam deflection as a function of position. A = 50Mpa, at 412kHz. The photo on the right shows the overall beam profile at 30cm from the lens. This is a Bessel beam profile.}
\label{fig:Bessel_beam_profile}
\end{figure}

Rather than obtaining as smooth gradient in the refractive, one finds that the maximum is at the center and there a decreasing periodic modulation in the refractive index, as suggested in Figure \ref{fig:Calculated_refractive_index_profile_in_glycerin}. These refractive index modulations can be observed by deflection of a laser beam. Figure \ref{fig:Bessel_beam_profile} shows the results for this experiment. 

Figure \ref{fig:Calculated_refractive_index_profile_in_glycerin} shows the calculated refractive index profile for glycerin with an assumed pressure of 200 MPa in the center.  The refractive index change is on the order of 0.1\%, but as shown in Figure \ref{fig:Bessel_beam_profile} these seemingly small changes in the refractive index are large enough to deflect the beam by over a millimeter. Based on these deflections, the pressure is determined to be about 50 MPa. Further, as shown in both Figure \ref{fig:Calculated_refractive_index_profile_in_glycerin} and \ref{fig:Bessel_beam_profile} there are radial modulations in the refractive index and these of course result in modulations of the light as shown in the beam-profile of Figure \ref{fig:Bessel_beam_profile}.

These circular refractive index modulations in the ring produce an overall global behavior known as an axicon lens. Axicon lenses are used to produce optical beams with a Bessel and Mathieu function profile (Figure \ref{fig:Bessel_beam_profile}). An axicon fit is shown in Figure \ref{fig:Calculated_refractive_index_profile_in_glycerin}. These beams are said to be self-healing and diffraction-free. The intensity of the profile does not diverge diffractively as a function of distance from the lens, and diffraction around objects is eliminated. The beam is self-healing because the intensity pattern re-establishes itself some distance away from an obstruction.

\begin{figure}
\centering
\includegraphics[height=\columnwidth,angle=270]{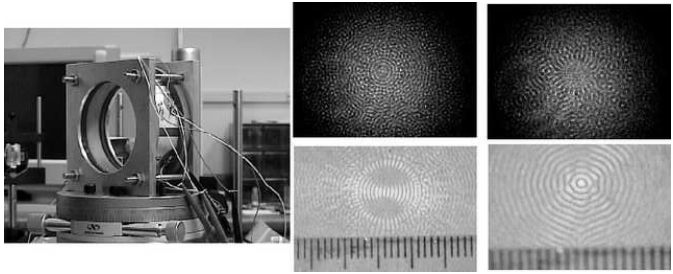}
\caption{The figure on the left shows an acoustic driven lens with sectioned transducers that can be driven independently. The top two photos show two of the many light patterns created and the lower two photos show two of the many standing wave pressure modulations possible in the fluid lens.}
\label{fig:fluid_lens}
\end{figure}
More recent work has used multitransducers and sectioned ring transducers to provide more degrees of freedom in the fluid modulations. Figure \ref{fig:fluid_lens} shows a photo of the device and several other photos. The bottom two photos were made by injecting 6 micron polystyrene spheres into the cell filled with water. The individual transducers were powered up separately, and by a process know as acoustophoresis the micro-scale particles will collect at the pressure nodes in the acoustic standing wave. A mm-scale has been placed on the lens for reference purposes. 

The upper two photos are examples of the light patterns generated on the output side of the lens while it is powered up at different frequencies. It is likely that an infinite number of possible image patterns can be generated from the equally infinite number of particle patterns in the fluid. It is possible, to drive each of the transducers at a different frequency, to amplitude or frequency modulate the drive frequency at any one of the transducers, and to drive the transducers in some predefined sequence to produce desired optical images at the output. Naturally, with this many degrees of freedom, we expect to use a genetic algorithm to evolve the transducer drive frequencies and patterns.

\section{Conclusions}

The results presented in this paper demonstrate that we can use bulk matter for computations almost as if it was programmable matter. With a programmable ``sea of gates'' in silicon we evolved an analogue oscillating circuit. In a liquid crystal substrate we evolved a tone discriminator circuit. In a magnetic dot substrate we evolved (in numerical simulations) a pattern matching system. In more speculative work, we have modelled a programmable Fermi surface where we manipulated the charge density wave in a crystal lattice. Finally, we outlined some laboratory work in using acoustic pulses to modulate bulk matter to affect the refractive index of light for using the ``constructed'' optical device in optical computing.

In each of the examples the computation comes about through local and nearest neighbour interactions, and the data and the program for the computations are fed directly into the bulk matter through the edges. If we had detailed models of the nano-scale physics, the computational burden would be significantly beyond foreseeable computer hardware. This can be circumvented through evolutionary programming. There are likely to be many more computational substrates that can be exploited through evolutionary engineering. 

\section*{Acknowledgment}
We thank Lornez Huelsbergen, Robert Slous, Brian Koen, Keith Higginson, Mike Costolo, and Oliver Rudolph for technical assistance on various stages of these projects. We further thank Bell Laboratories, Starlab – Brussels, and DARPA for funding on some of these subprojects.
\baselineskip0.7cm       
\bibliographystyle{IEEEtran} 
\bibliography{matphys}

\begin{thebibliography}{10}
\providecommand{\url}[1]{#1}
\csname url@rmstyle\endcsname
\providecommand{\newblock}{\relax}
\providecommand{\bibinfo}[2]{#2}
\providecommand\BIBentrySTDinterwordspacing{\spaceskip=0pt\relax}
\providecommand\BIBentryALTinterwordstretchfactor{4}
\providecommand\BIBentryALTinterwordspacing{\spaceskip=\fontdimen2\font plus
\BIBentryALTinterwordstretchfactor\fontdimen3\font minus
  \fontdimen4\font\relax}
\providecommand\BIBforeignlanguage[2]{{%
\expandafter\ifx\csname l@#1\endcsname\relax
\typeout{** WARNING: IEEEtran.bst: No hyphenation pattern has been}%
\typeout{** loaded for the language `#1'. Using the pattern for}%
\typeout{** the default language instead.}%
\else
\language=\csname l@#1\endcsname
\fi
#2}}

\bibitem{Wolfram:1985}
S.~Wolfram, ``Undecidability and intractability in theoretical physics,'' in
  \emph{Phys. Rev. Lett.}, vol. 54 (8), 1985, pp. 735--738.

\bibitem{Moore:1990}
C.~Moore, ``Unpredictability and undecidability in dynamical systems,'' in
  \emph{Phys. Rev. Lett}, vol.~64, 1990, pp. 2354--2357.

\bibitem{feynman67}
R.~P. Feynman, \emph{The Character of Physical Law}.\hskip 1em plus 0.5em minus
  0.4em\relax MIT Press, 1967.

\bibitem{Wolfram:2002}
S.~Wolfram, \emph{A New Kind of Science}.\hskip 1em plus 0.5em minus
  0.4em\relax Wolfram Media, 2002.

\bibitem{Yashihito:1994}
A.~Yashihito, ``Information processing using intelligent materials -
  information-processing architectures for material processors,'' \emph{J. of
  Intell. Mat. Syst. and Structures}, vol.~5, pp. 418--423, 1994.

\bibitem{Miller:2002}
J.~F. Miller and K.~Downing, ``Evolution in materio: Looking beyond the silicon
  box,'' in \emph{NASA/DOD Conference on Evolvable Hardware}.\hskip 1em plus
  0.5em minus 0.4em\relax IEEE Comp. Soc. Press, 2002, pp. 167--176.

\bibitem{harding04a}
S.~Harding and J.~F. Miller, ``Evolution in materio: Initial experiments with
  liquid crystal,'' in \emph{Proceedings of 2004 NASA/DoD Conference on
  Evolvable Hardware (EH'04)}, 2004, pp. 298--305.

\bibitem{harding04b}
------, ``Evolution in materio: A tone discriminator in liquid crystal,'' in
  \emph{In Proceedings of the Congress on Evolutionary Computation 2004
  (CEC'2004)}, vol.~2, 2004, pp. 1800--1807.

\bibitem{Wolfram:1986}
S.~Wolfram, ``Approaches to complexity engineering,'' in \emph{Physica D},
  vol.~22, 1986, pp. 385--399.

\bibitem{Thompson:1996}
A.~Thompson, I.~Harvey, and P.~Husbands, ``Unconstrained evolution and hard
  consequences,'' in \emph{Towards Evolvable Hardware, The Evolutionary
  Engineering Approach}, E.~Sanchez and M.~Tomassini, Eds.\hskip 1em plus 0.5em
  minus 0.4em\relax Springer, New York, NY, 1996, pp. 136--165.

\bibitem{Thompson:1997}
A.~Thompson, ``An evolved circuit, intrinsic in silicon, entwined with
  physics,'' in \emph{Evolvable Systems: From Biology to Hardware}, T.~Higuchi,
  M.~Iwata, and W.~Liu, Eds.\hskip 1em plus 0.5em minus 0.4em\relax Springer,
  New York, NY, 1997, pp. 390--405.

\bibitem{Thompson:2000}
A.~Thompson and C.~Wasshuber, ``Design of single electron systems through
  artificial evolution,'' in \emph{Int. J. Circuit Theory and Applications},
  vol. 28 (6), 2000, pp. 585--599.

\bibitem{Garvie:2003}
M.~Garvie and A.~Thompson, ``Evolution of self-diagnosing hardware,'' in
  \emph{Evolvable Systems, From Biology to Hardware}, Tyrrell, Haddow, and
  Torresen, Eds.\hskip 1em plus 0.5em minus 0.4em\relax Springer, 2003, pp.
  238--248.

\bibitem{Raichman:2003}
N.~Raichman, E.~Ben-Jacob, and R.~Segev, ``Evolvable hardware: Genetic search
  in a physical realm,'' in \emph{Physica A}, vol. 326, 2003, pp. 265--285.

\bibitem{Huelsbergen:1998}
L.~Huelsbergen, E.~A. Rietman, and R.~Slous, ``Evolution of astable
  multivibrators in silico,'' in \emph{Evolvable Systems: From Biology to
  Hardware}, M.~Sipper, D.~Mange, and A.~Perez-Uribe, Eds.\hskip 1em plus 0.5em
  minus 0.4em\relax Springer, New York, 1998, pp. 66--77.

\bibitem{Huelsbergen:1999}
------, ``Evolution of astable multivibrators in silico,'' in \emph{IEEE
  Transactions on Evoluationary Computing}, vol. 3 (3), 1999, pp. 197--204.

\bibitem{Thompson}
A.~Thompson, ``An evolved circuit, intrinsic in silicon, entwined with
  physics,'' in \emph{{ICES}}, 1996, pp. 390--405.

\bibitem{Miller02}
J.~F. Miller and K.~Downing, ``Evolution in materio: Looking beyond the silicon
  box,'' \emph{Proceedings of NASA/DoD Evolvable Hardware Workshop}, pp.
  167--176, July 2002.

\bibitem{Toffoli99}
T.~Toffoli, ``Programmable matter methods,'' \emph{Future Generation Computer
  Systems}, vol.~16, 1999.

\bibitem{Demus98}
D.~Demus, J.~Goodby, G.~W. Gray, H.~W. Spiess, and V.~Vill, Eds.,
  \emph{Handbook of Liquid Crystals}.\hskip 1em plus 0.5em minus 0.4em\relax
  Wiley-VCH, July 1998, vol. 1,2A,2B,3.

\bibitem{Khoo95}
I.~C. Khoo, \emph{Liquid Crystals: physical properties and nonlinear optical
  phenomena}.\hskip 1em plus 0.5em minus 0.4em\relax Wiley, 1995.

\bibitem{Khoo98}
I.-C. Khoo, S.~Slussarenko, B.~D. Guenther, M.-Y. Shih, P.~Chen, and W.~V.
  Wood, ``Optically induced space-charge fields, dc voltage, and
  extraordinarily large nonlinearity in dyedoped nematic liquid crystals,''
  \emph{Optics Letter}, vol.~23, no.~4, pp. 253--255, 1998.

\bibitem{Chandrasekhar98}
S.~Chandrasekhar, \emph{Handbook of Liquid Crystals}, D.~Demus, J.~Goodby,
  G.~W. Gray, H.~W. Spiess, and V.~Vill, Eds.\hskip 1em plus 0.5em minus
  0.4em\relax Wiley, 1998, vol.~2B.

\bibitem{Crossland98}
W.~A. Crossland and T.~D. Wilkinson, \emph{Nondisplay applications of liquid
  crystals}, D.~Demus, J.~Goodby, G.~W. Gray, H.~W. Spiess, and V.~Vill,
  Eds.\hskip 1em plus 0.5em minus 0.4em\relax Wiley, 1998, vol.~1.

\bibitem{naumov:98}
\BIBentryALTinterwordspacing
A.~F. Naumov, M.~Y. Loktev, I.~R. Guralnik, and G.~Vdovin, ``Liquid-crystal
  adaptive lenses with modal control,'' in \emph{Optics Letters}, vol.~23,
  1998, pp. 992--994. [Online]. Available:
  \url{http://www.okotech.com/mirrors/lens/}
\BIBentrySTDinterwordspacing

\bibitem{Layzell1998}
P.~Layzell, ``A new research tool for intrinsic hardware evolution,''
  \emph{Proceedings of The Second International Conference on Evolvable
  Systems: From Biology to Hardware, LNCS}, vol. 1478, pp. 47--56, 1998.

\bibitem{Crooks02}
J.~Crooks, ``Evolvable analogue hardware,'' Meng Project Report, The University
  Of York, 2002.

\bibitem{Calude:1998}
C.~S. Calude, J.~Casti, and M.~J. Dinneen, Eds., \emph{Unconventional Models of
  Computation}.\hskip 1em plus 0.5em minus 0.4em\relax Springer, New York, NY,
  1998.

\bibitem{GromB:1998}
T.~GromB, S.~Bornholdt, M.~GroB, M.~Mitchell, and T.~Pellizzari,
  \emph{Non-Standard Computation}.\hskip 1em plus 0.5em minus 0.4em\relax
  Wiley-VCH, New York, NY, 1998.

\bibitem{Siegelmann:1999}
H.~T. Siegelmann, \emph{Neural Networks and Analog Computation, Beyond the
  Turing Limits}.\hskip 1em plus 0.5em minus 0.4em\relax Birkhauser, Boston,
  MA, 1999.

\bibitem{Sipper:1997}
M.~Sipper, \emph{Evolution of Parallel Cellular Machines, The Cellular
  Programming Approach}.\hskip 1em plus 0.5em minus 0.4em\relax Springer, New
  York, NY, 1997.

\bibitem{Mange:1998}
D.~Mange and M.~Tomassini, Eds., \emph{Bio-Inspired Computing Machines}.\hskip
  1em plus 0.5em minus 0.4em\relax Presses Polytechniques et Universitaires
  Romandes, Switzerland, 1998.

\bibitem{Sienko:2003}
T.~Sienko, A.~Adamatzky, N.~Rambidi, and M.~Conrad, \emph{Molecular
  Computing}.\hskip 1em plus 0.5em minus 0.4em\relax MIT Press, Cambridge, MA,
  2003.

\bibitem{Benioff:1980}
P.~Benioff, ``The computer as a physical system: A microscopic quantum
  mechanical hamiltonian model of computers as represented by turing
  machines,'' in \emph{J. of Statistical Physics}, vol. 22 (5), 1980, pp.
  563--591.

\bibitem{Benioff:1982}
------, ``Quantum mechanical hamiltonian models of turing machines,'' in
  \emph{J. of Statistical Physics}, vol. 29 (3), 1982, pp. 515--546.

\bibitem{Albert:1983}
D.~Z. Albert, ``On quantum mechanical automata,'' \emph{Physics Letters}, vol.
  98A (5,6), pp. 249--252, 1983.

\bibitem{Biafore:1994}
M.~Biafore, ``Cellular automata for nanometer-scale computation,'' in
  \emph{Physica}, vol. D 70, 1994, pp. 415--433.

\bibitem{Porod:1999}
W.~Porod, C.~S. Lent, G.~H. Bernstein, A.~O. Orlov, I.~Amlani, G.~L. Snider,
  and J.~L. Merz, ``Quantum-dot cellular automata: Computing with coupled
  quantum dots,'' in \emph{Int. J. Electronics}, vol. 86 (5), 1999, pp.
  549--590.

\bibitem{Penrose:1989}
R.~Penrose, \emph{The Emperor's New Mind, Concerning Computers, Minds, and the
  Laws of Physics}.\hskip 1em plus 0.5em minus 0.4em\relax Oxford University,
  Oxford, UK, 1989.

\bibitem{Penrose:1994}
------, \emph{Shadows of the Mind, A Search for the Missing Science of
  Consciousness}.\hskip 1em plus 0.5em minus 0.4em\relax Oxford University
  Press, Oxford, UK,, 1994.

\bibitem{Satinover:2001}
J.~Satinover, \emph{The Quantum Brain}.\hskip 1em plus 0.5em minus 0.4em\relax
  John Wiley, New York, NY, 2001.

\bibitem{Kak:1992}
S.~C. Kak, ``State generators and complex neural memories,'' \emph{Pramana, J.
  of Physics}, vol. 38 (3),, pp. 271--278, 1992.

\bibitem{Kak:1995}
------, ``On quantum neural computing,'' \emph{Information Sciences}, vol.~83,
  pp. 143--160, 1995.

\bibitem{Ventura:1999}
D.~Ventura, ``Implementing competitive learning in a quantum system,'' in
  \emph{Proceedings of IJCNN}, vol. CD Version, 1999.

\bibitem{Behrman:1999}
E.~C. Behrman, J.~E. Steck, and S.~R. Skinner, ``A spatial quantum neural
  computer,'' in \emph{Proceedings of IJCNN}, vol. CD Version,, 1999.

\bibitem{Zak:1999}
M.~Zak, ``Quantum analog computing,'' in \emph{Chaos, Solitons and Fractals},
  vol. 10 (10), 1999, pp. 1583--1620.

\bibitem{Mahler:1995}
G.~Mahler and V.~A. Weberruss, \emph{Quantum Networks, Dynamics of Open
  Nanostructures}.\hskip 1em plus 0.5em minus 0.4em\relax Springer, New York,
  NY, 1995.

\bibitem{Smith:1990}
H.~I. Smith and D.~A. Antoniadis, ``Seeking a radically new electronics,'' in
  \emph{MIT Technology Review,}, vol. April, 1990, pp. 27--40.

\bibitem{Chakraborty:1999}
T.~Chakraborty, \emph{Quantum Dots, A Survey of the Properties of Artificial
  Atoms}.\hskip 1em plus 0.5em minus 0.4em\relax North-Holland, Amsterdam,
  1999.

\bibitem{Zorpette:2001}
G.~Zorpette, ``The quest for the spin transistor,'' in \emph{IEEE Spectrum},
  vol. 38 (12), 2001, pp. 30--35.

\bibitem{DasSarma:2001}
S.~D. Sarma, ``Spintronics,'' in \emph{American Scientist}, vol. 89 (6), 2001,
  pp. 516--523.

\bibitem{Wolf:2001}
S.~A. Wolf and et~al, ``Spintronics: A spin-based electronics vision for the
  future,'' in \emph{Science}, vol. 294, 2001, pp. 1488--1495.

\bibitem{Kirczenow:1993}
G.~Kirczenow, B.~L. Johnson, J.~C. Barnes, and R.~Akis, ``Novel quantum hall
  phenomena in arrays of quantum dots,'' in \emph{NanoStructured Materials},
  vol.~3, 1993, pp. 125--135.

\bibitem{Matsueda:1999}
H.~Matsueda, ``Spatiotemporal dynamics of quantum computing dipole-dipole block
  systems,'' in \emph{Chaos, Solitons and Fractals}, vol. 10 (10), 1999, pp.
  1737--1748.

\bibitem{Cowburn:1999}
R.~P. Cowburn, D.~K. Koltsov, A.~O. Adeyeye, and M.~E. Welland, ``Single-domain
  circular nanomagnetics,'' in \emph{Phys. Rev. Lett}, vol. 83 (5), 1999, pp.
  1042--1045.

\bibitem{Yu:2002}
C.~Yu, J.~Pearson, and D.~Li, ``Magnetic domains and magnetostatic interactions
  of self-assembled co dots,'' in \emph{J. of Appl. Phys}, vol. 91 (10, 2002,
  pp. 6955--6957.

\bibitem{Bar-Yam:1997}
Y.~Bar-Yam, \emph{Dynamics of Complex Systems}.\hskip 1em plus 0.5em minus
  0.4em\relax Addison-Wesley, Reading, MA, 1997.

\bibitem{Mezard:1986}
M.~Mezard, ``On the statistical physics of spin glasses,'' in \emph{Disordered
  Systems and Biological Organization}, E.~Bienenstock, F.~F. Soulie, and
  G.~Weisbuch, Eds.\hskip 1em plus 0.5em minus 0.4em\relax Springer-Verlag,
  1986, pp. 19--132,.

\bibitem{Hayes:2000}
B.~Hayes, ``The world in a spin,'' in \emph{American Scientist}, vol.~88, 2000,
  pp. 384--388.

\bibitem{Hopfield:1982}
J.~J. Hopfield, ``Neural networks and physical systems with emergent collective
  computational abilities,'' in \emph{Proceedings of the National Academy of
  Sciences}, vol. 79,, 1982, pp. 2554--2558.

\bibitem{Bienenstock:1986}
E.~Bienenstock, F.~F. Soulie, and G.~Weisbuch, \emph{Disordered Systems and
  Biological Organization}.\hskip 1em plus 0.5em minus 0.4em\relax
  Springer-Verlag, 1986.

\bibitem{Landauer:1995}
R.~Landauer, ``Is quantum mechanics useful?'' in \emph{Phil. Trans. R. Soc.
  London}, vol. A 353, 1995, pp. 367--376.

\bibitem{Roychowdhury:1997}
V.~P. Roychowdhury, D.~B. Janes, and S.~Bandyopadhyay, ``Nanoelectronic
  architecture for boolean logic,'' in \emph{Proceedings of the IEEE}, vol. 85
  (4), 1997, pp. 574--587.

\bibitem{Goldberg89}
D.~Goldberg, \emph{Genetic Algorithms in Search, Optimization and Machine
  Learning}.\hskip 1em plus 0.5em minus 0.4em\relax Reading, Massachusetts:
  Addison-Wesley, 1989.

\bibitem{Shinjo:2000}
T.~Shinjo, T.~Okuno, R.~Hassdorf, K.~Shigeto, and T.~Ono, ``Magnetic vortex
  core observation in circular dots of permalloy,'' in \emph{Science}, vol.
  289, 2000, pp. 930--932.

\bibitem{Cowburn:2000}
R.~P. Cowburn and M.~E. Welland, ``Room temperature magnetic quantum cellular
  automata,'' in \emph{Science}, vol. 287, 2000, pp. 1466--1468.

\bibitem{Prinz:1998}
G.~A. Prinz, ``Magnetoelectronics,'' in \emph{Science}, vol. 282, 1998, pp.
  1660--1663.

\bibitem{Gorelik:2003}
L.~Y. Gorelik, R.~I. Shekhter, V.~M. Vinokur, D.~E. Feldman, V.~I. Kozub, and
  M.~Jonson, ``Electrical manipulation of nanomagnets,'' in \emph{Physical Rev.
  Lett}, vol. 91 (8), 2003, p. 088301.

\bibitem{Hassoun:1995}
M.~H. Hassoun.\hskip 1em plus 0.5em minus 0.4em\relax MIT Press, Cambridge, MA,
  1995.

\bibitem{Rumelhart:1986}
D.~E. rumelhart, J.~L. McClelland, and the PDP Research~Group, \emph{Parallel
  Distributed Processing}.\hskip 1em plus 0.5em minus 0.4em\relax MIT Press,
  Cambridge, MA, 1986.

\bibitem{Wilson:1975}
J.~A. Wilson, F.~J.~D. Salvo, and S.~Mahajan, ``Charge-density waves and
  superlattices in the metallic layered transition metal dichalcogenides,'' in
  \emph{Advances in Physics}, vol.~24, 1975, p. 117.

\bibitem{Gruner:1994}
G.~L. Gruner, ``Density waves in solids.''\hskip 1em plus 0.5em minus
  0.4em\relax Addison-Wesley, New York, 1994.

\bibitem{Gill:1986}
J.~C. Gill, ``Thermally initiated phase-slip in the motion and relaxation of
  charge density waves in niobium triselenide,'' in \emph{J. Phys. C: Solid
  State Physics}, vol.~19, 1986, pp. 6589--6604.

\bibitem{Fukuyama:1978}
H.~Fukuyama and P.~A. Lee, ``Dynamics of charge density wave. i. impurity
  pinning in a single chain,'' in \emph{Phys. Rev. B}, vol.~17, 1978, pp.
  535--541.

\bibitem{Azbel:1984}
M.~Azbel and P.~Bak, ``Analytical results on the periodically driven damped
  pendulum. applications to sliding charge-density waves and josephson
  junctions,'' in \emph{Physical Rev. B}, vol.~30, 1984, pp. 3722--3727.

\bibitem{Zettl:1986}
A.~Zettl, M.~S. Sherwin, and R.~P. Hall, ``Dynamics of charge density wave
  conductors: Broken coherence, chaos, and noisy precursors,'' in \emph{Physica
  B}, vol. 143, 1986, pp. 69--72.

\bibitem{Sherwin:1988}
M.~S. Sherwin, A.~Zettl, and R.~P. Hall, ``Switching and charge-density wave
  transport in nbse3. iii. dynamical instabiilties,'' in \emph{Phys. Rev. b},
  vol.~38, 1988, pp. 13\,028--13\,046.

\bibitem{Inui:1988}
M.~Inui, R.~P. Hall, S.~Doniach, and A.~Zettl, ``Phase slips and switching in
  charge density wave transport,'' in \emph{Physical Rev. B}, vol.~33, 1988,
  pp. 13\,047--13\,059.

\bibitem{Cava:1984}
R.~J. Cava, R.~M. Fleming, P.~Littlewood, E.~A. Rietman, L.~F. Schneemeyer, and
  R.~G. Dunn, ``Dielectric response of the charge-density wave in {${\rm
  K}_{{\rm 0}{\rm .3}} {\rm MoO}_{\rm 3}$},'' in \emph{Physical Rev}, vol. B 30
  (6), 1984, pp. 3228--3239.

\bibitem{Cava:1985}
R.~J. Cava, L.~F. Schneemeyer, R.~M. Fleming, P.~B. Littlewood, and E.~A.
  Rietman, ``Effect of impurities on the dielectric response of the
  charge-density wave in {${\rm K}_{{\rm 0}{\rm .3}} {\rm MoO}_{\rm 3}$},'' in
  \emph{Physical Rev}, vol. B 32 (6), 1985, pp. 4088--4096.

\bibitem{Csiba:1989}
T.~Csiba, G.~Kriza, and A.~Janossy, ``Charge density wave noise propagation in
  the blue bronzes {${\rm Rb}_{{\rm 0}{\rm .3}} {\rm MoO}_{\rm 3}$} and {${\rm
  K}_{{\rm 0}{\rm .3}} {\rm MoO}_{\rm 3}$},'' in \emph{Physical Rev}, vol. B 40
  (15), 1989, pp. 10\,088--10\,099.

\bibitem{Karttunen:1999}
M.~Kattuneu, M.~Haataja, K.~R. Elder, and M.~Grant, ``Defects, order, and
  hysteresis in driven charge-density waves,'' in \emph{Phys. Rev. Lett.},
  vol.~83, 1999, pp. 3518--3521.

\bibitem{Hall:1988a}
R.~P. Hall, M.~F. Hundley, and A.~Zettl, ``Switching and charge-density wave
  transport in {${\rm NbSe}_3$}. i. dc characteristics,'' in \emph{Physical
  Rev. B}, vol.~38, 1988, pp. 13\,002--13\,018.

\bibitem{Hall:1988b}
R.~P. Hall and A.~Zettl, ``Switching and charge-density wave transport in
  {${\rm NbSe}_3$}. ii. ac characteristics,'' in \emph{Phys. Rev. B}, vol.~38,
  1988, pp. 13\,019--13\,027.

\bibitem{Myers:1993}
C.~R. Myers and J.~P. Sethna, ``Collective dynamics in a model of sliding
  charge-density waves. i. critical behavior,'' in \emph{Physical Rev}, vol. B,
  47, 1993, pp. 11\,171--11\,193.

\bibitem{Kanek:1991}
Kanek, ``Globally coupled circle maps,'' in \emph{Physica}, vol. D 54, 1991,
  pp. 5--19.

\bibitem{Sinha:1998}
S.~Sinha and W.~L. Ditto, ``Dynamics based computation,'' in \emph{Physical
  Rev. Lett}, vol.~81, 1998, pp. 2156--2159.

\bibitem{Sinha:1999}
------, ``Computing with distributed chaos,'' in \emph{Physical Rev}, vol.~E,
  1999, pp. 363--377.

\bibitem{Sinha:2002a}
S.~Sinha, T.~Munakata, and W.~L. Ditto, ``Parallel computing with extended
  dynamical systems,'' in \emph{Physical Rev}, vol. E, 65, 036214, 2002.

\bibitem{Sinha:2002b}
------, ``Flexible parallel implementation of logic gates using chaotic
  elements,'' in \emph{Physical Rev}, vol. E, 65, 036216, 2002.

\bibitem{Rietman98}
E.~A. Rietman, \emph{Experiments in Artificial Neural Networks}.\hskip 1em plus
  0.5em minus 0.4em\relax Tab Books, 1998.

\bibitem{Mantel:1997}
O.~C. Mantel, H.~S.~H. van~der Zant, A.~J. Steinfort, and C.~Dekker, ``Thin
  films of the charge-density-wave oxide rb0.3moo3 by pulsed-laser
  deposition,'' in \emph{Phys. Rev. B}, vol. 55 (7), 1997), pp. 4817--4824.

\bibitem{Zant:1997}
H.~S.~J. van~der Zant, O.~C. Mantel, C.~P. Heij, and C.~Dekker,
  ``Photolographic patterning of the charge-density-wave conductor rb0.3moo3,''
  in \emph{Synthetic Metals}, vol.~86, 1997, pp. 1781--1784.

\bibitem{Higginson:2004a}
K.~A. Higginson, M.~A. Costolo, and E.~A. Rietman, ``Adaptive geometric optics
  derived from nonlinear acoustic effects,'' in \emph{Appl. Phys. Lett.}, vol.
  84 (6), 2004, pp. 843--845.

\bibitem{Higginson:2004b}
K.~A. Higginson, M.~A. Costolo, E.~A. Rietman, B.~Lipkens, and J.~M. Ritter,
  ``Tunable optics derived from nonlinear acoustic effects,'' vol. 95 (10),
  2004.

\bibitem{Brillouin:1922}
Brillouin:1922, ``Diffusion de la lumiere et des rayons-x par un corps
  transparent homogene: influence de l'agitation termique,'' \emph{Ann. Phys.
  (Paris)}, vol.~17, pp. 88--122, 1922.

\bibitem{Deby:1932}
P.~Debye and F.~W. Sears, ``On the scattering of light by supersonic waves,''
  in \emph{Proc. Natl. Acad. Sci. USA}, vol.~18, 1932, pp. 409--414.

\bibitem{Korpel:1966}
A.~Korpel, R.~Adler, P.~Desmares, and W.~Watson, ``A television display using
  acoustic deflection and modulation of coherent light,'' in \emph{Applied
  Optics}, vol. 5 (10), 1966, pp. 1667--1675.

\bibitem{Hamilton:1998}
M.~F. Hamilton and D.~T. Blackstock, Eds., \emph{Nonlinear Acoustics}.\hskip
  1em plus 0.5em minus 0.4em\relax Academic Press, New York, 1998.

\end{thebibliography}
\newpage 
\begin{biographynophoto}{Dr. Julian F. Miller}
Julian F. Miller obtained a BSc in Physics at the University of London in 1980. He obtained a PhD in Mathematics at the City University in 1988. He obtained a Postgraduate Certificate in Teaching and Learning in Higher Education at the University of Birmingham in 2002. He is currently a lecturer in the Department of Electronics at the University of York.  His research interests are: genetic programming, evolvable hardware, artificial life and quantum computing. Dr. Miller is an Associate Editor of the IEEE Transactions on Evolutionary Computation, Associate Editor of Genetic Programming and Evolvable Machines, and editor board member of the journals Evolutionary Computation and Unconventional Computing. He has chaired various conferences in the fields of Genetic Programming and Evolvable Hardware.
\end{biographynophoto}

\begin{biographynophoto}{Dr. Edward A. Reitman}
Edward A. Rietman has B.S. degrees in physics and chemistry, a B.A. degree in philosophy, an M.S. in materials science, a Ph.D. in physics, and a graduate degree in bioinformatics. He spent 19 years at Bell Labs, where he worked on solid-state physics, neural network hardware and AI applications for control of CMOS manufacturing. He has published several books on neural networks, chaos, parallel computing, artificial life and nanotechnology. In addition he has authored (or coauthored) over 100 technical papers and dozens of patents. He is currently doing research in applied physics and AI applications. He is a member of the IEEE, APS, and AAAS.
\end{biographynophoto}


\begin{biographynophoto}{Simon Harding}
Simon Harding has PhD in Electronic Engineering (University of York, 2005) and a B.Sc. degree in artificial intelligence and computer science (University of Birmingham UK 2001).   His research interests include evolvable hardware, genetic programming, developmental systems and self organising systems. He is currently researching at Memorial University.
\end{biographynophoto}

\end{document}